# Order evolution from a high-entropy matrix: understanding and predicting paths to low temperature equilibrium

Saeed S. I. Almishal[1], Leixin Miao[1], Yueze Tan[1], George N. Kotsonis[1], Jacob T. Sivak[2], Nasim Alem[1], Long-Qing Chen[1], Vincent H. Crespi[1,3], Ismaila Dabo[1], Christina M. Rost,[4] Susan B. Sinnott[1,2], and Jon-Paul Maria[1]

[1]Department of Materials Science and Engineering, The Pennsylvania State University, University Park, PA 16802, USA
[2]Department of Chemistry, The Pennsylvania State University, University Park, PA 16802, USA
[3]Department of Physics, The Pennsylvania State University, University Park, PA 16802, USA
[4]Department of Materials Science and Engineering, Virginia Polytechnic Institute and State University, Blacksburg, VA 24060, USA

**Corresponding author:** Jon-Paul Maria and Saeed S. I. Almishal saeedsialmishal@gmail.com

## Abstract

Interest in high-entropy inorganic compounds originates from their ability to stabilize cations and anions in local environments that rarely occur at standard temperature and pressure. This leads to new crystalline phases in many-cation formulations with structures and properties that depart from conventional trends. The highest-entropy homogeneous and random solid-solution is a parent structure from which a continuum of lower-entropy offspring can originate by adopting chemical and/or structural order. This report demonstrates how synthesis conditions, thermal history, and elastic and chemical boundary conditions conspire to regulate this process in $Mg_{0.2}Co_{0.2}Ni_{0.2}Cu_{0.2}Zn_{0.2}O$, during which coherent CuO nano-tweeds and spinel nano-cuboids evolve. We do so by combining structured synthesis routes, atomic-resolution microscopy and spectroscopy, density functional theory, and a phase field modeling framework that accurately predicts the emergent structure and local chemistry. This establishes a framework to appreciate, understand, and predict the macrostate spectrum available to a high-entropy system that is critical to rationalize property engineering opportunities.

# Main

Solid solutions comprise a formulation space from which cornerstone material technologies spanning mechanical, optical, electronic, electromechanical and memory applications routinely emerge[1–3]. Examples of note include high strength-to-weight ratio alloys based on Al-Si[4], high-speed microcircuits on a Si-Ge platform[5], solid-state lasers built from $Al_{1-x}Ga_xN$ heterostructures[6], extreme-strain actuators from $(PbMg_{1/3}Nb_{2/3}O_3)_{1-x}(PbTiO_3)_x$ single crystals[7], and recently, Si-compatible ferroelectric thin films of $Al_{1-x}Sc_xN$[8] and $Al_{1-x}B_xN$[9]. The value afforded by these systems is the overarching ability to continuously tune composition between end members and subsequently vary structures and properties either smoothly between the two compositional bounds (as in GaAs-AlAs)[10], or through often-surprising maxima (as in $PbTiO_3$-$PbZrO_3$)[11]. Both categories yield rich science, property, and performance rewards.

Historically, the solid solution selection process uses isostructural end members, or end members whose elemental constituents share reasonably similar electronegativities, radii, and/or polyhedral coordinations to a common cation or anion. Classical criteria such as Goldschmidt's tolerance factor[12], Hume-Rothery rules[13], and Pauling's rules[14] provide excellent frameworks that guide selection. These frameworks favor structures with a large negative formation enthalpy, thus the compositional solubility spectrum they predict will directly reflect the dispersity of elemental periodic properties[15,16]. That is to say, only a finite number of enthalpy-driven combinations exist that afford extensive solubility windows because elemental properties vary rapidly along most directions of the periodic table[3,16].

Initiated by interest in amorphous metals, researchers in the 1990s began exploring formulation spaces involving five or more components with comparable mole fractions in a deliberate effort to maximize configurational disorder through chemistry[1,3]. Many diverse metal atoms on a single sublattice will boost configurational entropy and, with increasing temperature, will thermodynamically stabilize the single-phase solid solution. The pioneering works of Cantor[17] and Yeh[18] exemplify this approach. The same possibility was demonstrated by Rost *et al.* in 2015, with direct proof of entropic stabilization in complex oxides[19]. Comparable reports for carbides, borides, sulfides, and nitrides followed soon thereafter[3]. There are now many reports for high-entropy compositions that solubilize metal cations in unusual local coordinations and structural distortions that are associated with interesting property responses; examples include an energy landscape promoting ion migration or ion valence distributions that tune the electronic band gap[3,20,21].

These observations lead to an alternative perspective for high configurational entropy materials, one considering them as solid solutions in a composition landscape with boundaries well beyond the Hume-Rothery or Pauling limits. With this high configurational entropy by formulation, the the equilibrium temperature between enthalpic or entropic victory is within or close to practical synthesis limits, *i.e.*, before problematic melting, volatilization, or metal reduction. Consequently, the highest-entropy "parent" solid solution – that features a homogeneous and random elemental distribution – can be preserved at room temperature by quenching into a metastable state. This disordered parent is linked thermodynamically to a continuum of offspring sub-structures that are accessible through engineered thermal histories. Each offspring represents the outcome of an entropy-enthalpy exchange to an increasingly ordered state. This perspective is fundamentally the same one used to understand phase evolution in metal alloys[22], however, the higher diversity of metal cations in a high-entropy oxide (HEO) offers more degrees of freedom during the return to thermal equilibrium. Stated alternatively, such materials have more ways to spend their entropy.

We focus on the original high-entropy and entropy-stabilized oxide, $Mg_{0.2}Co_{0.2}Ni_{0.2}Cu_{0.2}Zn_{0.2}O$ that adopts an equilibrium rock salt solid solution structure above ~875 °C with all cations in octahedral coordination and no evidence of intermediate or long-range elemental clustering[19,23]. For all HEOs and $Mg_{0.2}Co_{0.2}Ni_{0.2}Cu_{0.2}Zn_{0.2}O$ bulk ceramics, some degree of quenching is needed to retain a locked metastable room temperature configurational microstate representative of the solid solution macrostate[3,24,25]. In our experience, normal furnace cooling at ~10 °C/min is sufficiently rapid, and years of dwell time at room temperature does not produce measurable structural changes. However, phase decomposition studies on bulk samples demonstrate that intermediate temperatures (roughly 300-850 °C) result in CuO precipitation within several hours[26,27]. This could be due to Cu-Cu pair interactions that reduce the total energy as they create new and unique lower-configurational-entropy macrostates[28,29].

We explore the continuum of new states that emerge when a metastable high-entropy solid solution approaches equilibrium upon "fluid" exchange between entropic disorder and enthalpic return. Note that the enthalpic return is bigger than the entropic spending, a condition required of any system that undergoes total free energy density reduction. We focus specifically on the early stages when new entropic motifs adopt order but sustain lattice coherence. Fig. 1 qualitatively illustrates a hypothetical HEO offspring spectrum. The right end member (Fig. 1d) represents the most entropy-rich state where 5 cations randomly occupy a single sublattice with no regular chemical ordering, while the left end member (Fig. 1a) represents a much lower-entropy state featuring chemical clustering and multiple phases. The region of present interest is the center, where more subtle hypothetical entropy spending occurs to chemically segregate a particular species (Fig. 1b) or to locally induce point defects involving specific cation species (Fig. 1c). Importantly, for the latter two examples, the anion sublattice is unchanged and the boundaries to new, partially-ordered structures are coherent and defect-free.

Interest in these still-coherent macrostates is predicated on an expectation that local distortions associated with phase instability will lead to new property possibilities. As an example, magnetic and electron-correlated properties are sensitive to highly localized exchange interactions that may be enhanced by entropy-reducing reconfigurations[30–32]. Another example occurs when short-range ordering impacts electrochemical utility to limit battery recharge cycles[33] or to facilitate reversible oxygen uptake and rejection cycles for high efficiency water splitting[34]. Inspired by these examples, we develop a predictive understanding for evolving order in high-entropy materials from a quenched metastable state, and report how offspring macrostates, featuring partially-ordered coherent nanostructures, emerge at the nanometer scale.

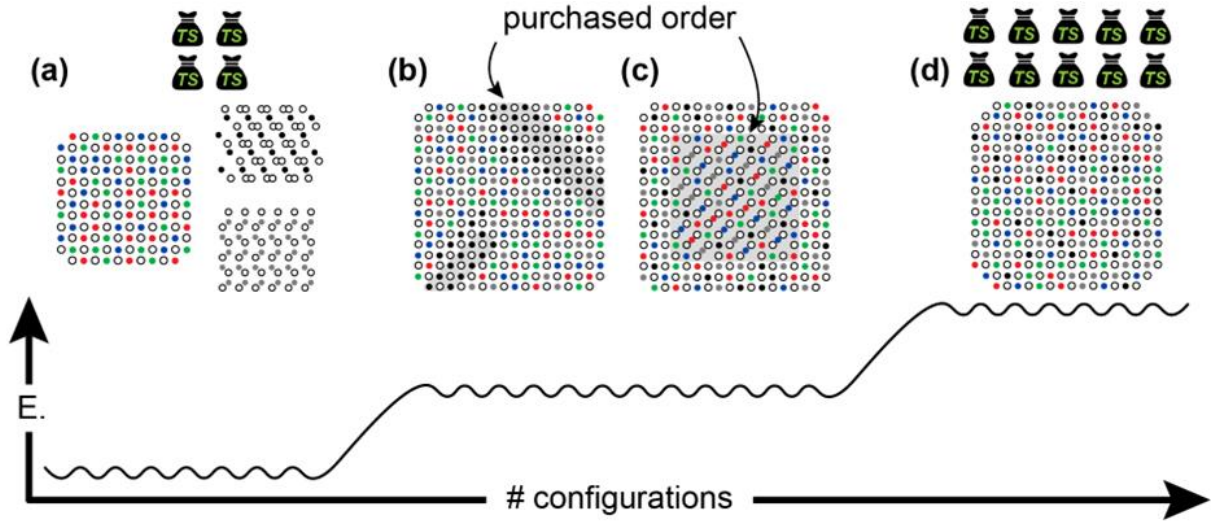

**Fig. 1:** Qualitative depiction of possible atomic configurations in a rock salt HEO different in total energy, E, plotted along one hypothetical pathway through configuration space. (a) represents a low-configurational-entropy state, including a three component rock salt solid solution and two other phases with different crystal structures each with a single cation; (b) a middle region with intermediate entropy featuring chemical segregation of a particular species concentrated nanoregions; (c) a second intermediate region featuring nanoregions where some of octahedral cations shift to tetrahedral interstitial sites without chemical ordering; and (d) the most entropy-rich state where 5 cations randomly occupy a single sublattice with no regular chemical ordering.

**The experiment: Exploring kinetic pathways to order during synthesis**

In principle, a specimen's atomic configuration and physical properties depend on synthesis conditions, which vary widely in kinetic terms[3]. Physical vapor deposition (PVD) techniques such as pulsed-laser deposition (PLD) produce gas phase particle ensembles with eV-level kinetic energies ($10^4$-$10^5$ K effective temperature) that equilibrate to $k_B T_{substrate}$ conditions just nanoseconds after substrate landing, which can secure a metastable, but persistent high-energy structure[3,35]. PLD is particularly well suited to explore these macrostate possibilities by establishing pulse rate, laser energy density, substrate temperature, and deposition pressure combinations that, for example, change average growth rate dramatically, while leaving adatom landing energy unchanged. In general pulse rate controls the time interval between adlayer accumulation where comparatively rapid surface diffusion mechanisms remain active, substrate temperature governs thermal vibrational entropy, film thickness regulates epitaxial strain accumulation, and pressure can regulate kinetic energy thermalization. Using such capabilities, we demonstrate nanostructure engineering within $Mg_{0.2}Co_{0.2}Ni_{0.2}Cu_{0.2}Zn_{0.2}O$ arising from the growing film's desire to approach equilibrium under these different mobility and driving force conditions.

We explore nanostructure evolution through entropy-enthalpy exchange in epitaxial $Mg_{0.2}Co_{0.2}Ni_{0.2}Cu_{0.2}Zn_{0.2}O$ thin films grown with PLD on warm substrates (~400°C). The laser repetition rate, which controls the average accumulation rate of material, varied between 1 and 5 Hz, and the deposition duration, which is quantified by the total number of laser pulses, varied between 1200 and 6000 pulses. Using these conditions, the following film categories are created:

**i.** thin and fast; **ii.** thin and slow; **iii.** thick and fast; and **iv.** thick and slow. Films from the four categories are denoted as 1200-5Hz, 1200-1Hz, 6000-5Hz, and 6000-1Hz, respectively. In this series, thin vs. thick delineates materials with different quantities of epitaxial and thermal expansion mismatch induced stored elastic strain energy, while slow vs. fast delineates materials with more or less time available for "annealing/relaxation" between flux pulse events.

X-ray diffraction scans provide the initial structural information for films prepared using these conditions. Fig. 2(a) shows the 1200-5Hz, 1200-1Hz, and 6000-1Hz films exhibit single-phase rock salt structure with pronounced thickness fringes suggesting high crystalline fidelity. The 6000-5Hz film exhibits an additional peak suggesting an additional phase. Reciprocal space maps of the (204) reflections in Fig. 2(b-e) indicate that all films are epitaxially strained with the same in-plane lattice parameter as the MgO substrate (4.21 Å). Calculations from out-of-plane scans reveal out of plane lattice constant values of 4.27 Å (1200-5Hz), 4.28 Å (1200-1Hz), 4.27 Å and 4.25 Å (6000-5Hz), and 4.27 Å (6000-1Hz). The laboratory XRD analysis indicates noteworthy structural changes for the different deposition conditions, despite the same overall deposition flux and the same substrate temperature.

Scanning transmisstion electron microscopy is used to explore structural features and local chemistry at the atomic scale. This technique is critical because, especially at the onset of ordering, longer range probes like diffraction will have great difficulty resolving structural changes. Representative samples from the four catagories were examined. The 1200-5Hz film SAED pattern shows sharp diffraction spots indicating a pristine rock salt crystal (Fig. 2f). The ADF-STEM micrograph appears featureless, and the higher magnification inset shows an abrupt epitaxial interface and clearly defined rock salt lattice (Fig. 2i). The 1200-1Hz film LAADF-STEM micrograph shows high-contrast striations indicative of local lattice strain aligned to the {101} planes in real space (Fig. 2j) and the SAED pattern (Fig. 2g) shows streaking in all <101> reciprocal lattice directions which is a "tweed" nanostructure signature. This tweed nanostructure is often reported in cubic alloy systems due to the formation of tetragonally distorted platelets[36,37]. The 6000-5Hz film LAADF-STEM mode (Fig. 2k) reveals starkly different features: a high density of nanoscopic cuboids that distribute randomly throughout the rock salt matrix **(supplementary Note 1)**. The 6000-5Hz film SAED pattern (Fig. 2h) exhibits extra spots forbidden in rock salt crystals, appearing at $\vec{g} = \{101\}$ (Fig. 2h inset). The alignment of these spots relative to the matrix indicates an epitaxial orientational arrangement in three dimensions.

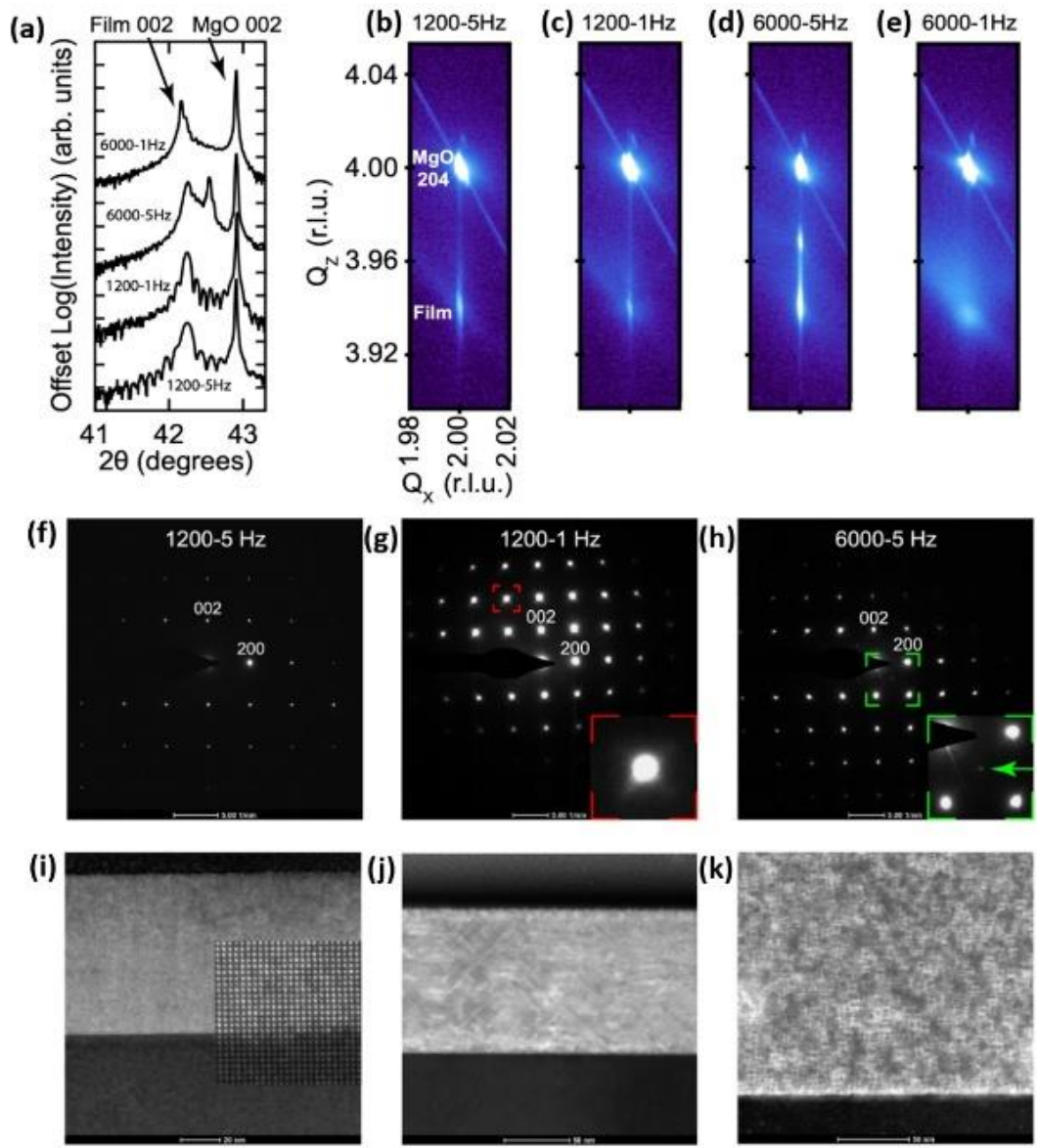

**Fig. 2:** Out-of-plane XRD patterns (a) and reciprocal space maps (b-e) of the four film categories. 1200-5Hz, 1200-1Hz, and 5000-5Hz exhibit single-phase rocks salt structure with pronounced thickness fringes and high crystalline fidelity. 6000-5Hz exhibits an additional peak. Reciprocal space maps of the (204) reflections indicate that 1200-5Hz, 1200-1Hz, and 6000-5Hz are epitaxially strained with the same in-plane lattice parameter as the MgO substrate. TEM analysis showing (f-h) SAED patterns 1200-5Hz, 1200-1Hz, and 6000-5Hz with insets highlighting the streaking and additional diffraction spots (acquired from the [010] zone axis); and (i-k) ADF STEM images for the same samples showing uniform structure in 1200-5Hz and nanostructure development in 1200-1Hz (nano-tweed morphology) and 6000-5Hz (nano-cuboid morphology). (j,k) were collected in LAADF-STEM mode to highlight strain contrast.

**Local symmetry and chemical analysis: (a) CuO-rich nanotweeds**

Fig. 3(a) presents a magnified ADF-STEM image of the 1200-1Hz film, showing the atomically resolved morphology of the nanoscale tweed microstructure. EDX mapping (Fig. 3b) reveals an enrichment of Cu ions and deficiency of Mg, Co, Ni, and Zn ions within the platelets. Unsupervised machine learning-based local symmetry analysis **(Supplementary Note 2)** was employed to categorize the local Fourier transform patterns collected from Fig. 3a into three classes: T1, a CuO-rich zone with an orientation pointing down and right; T2, a CuO-rich zone with an orthogonal orientation to T1; and RS, the rock salt matrix with no nanostructuring (Fig. 3C). Further analysis of T1 and T2 reveals lattice rotation and tetragonal distortions consistent with observations of Jahn-Teller distorted $CuO_6$ octahedra in $Mg_{0.2}Co_{0.2}Ni_{0.2}Cu_{0.2}Zn_{0.2}O$[23,38,39] **(described in detail in Supplementary Note 2).**

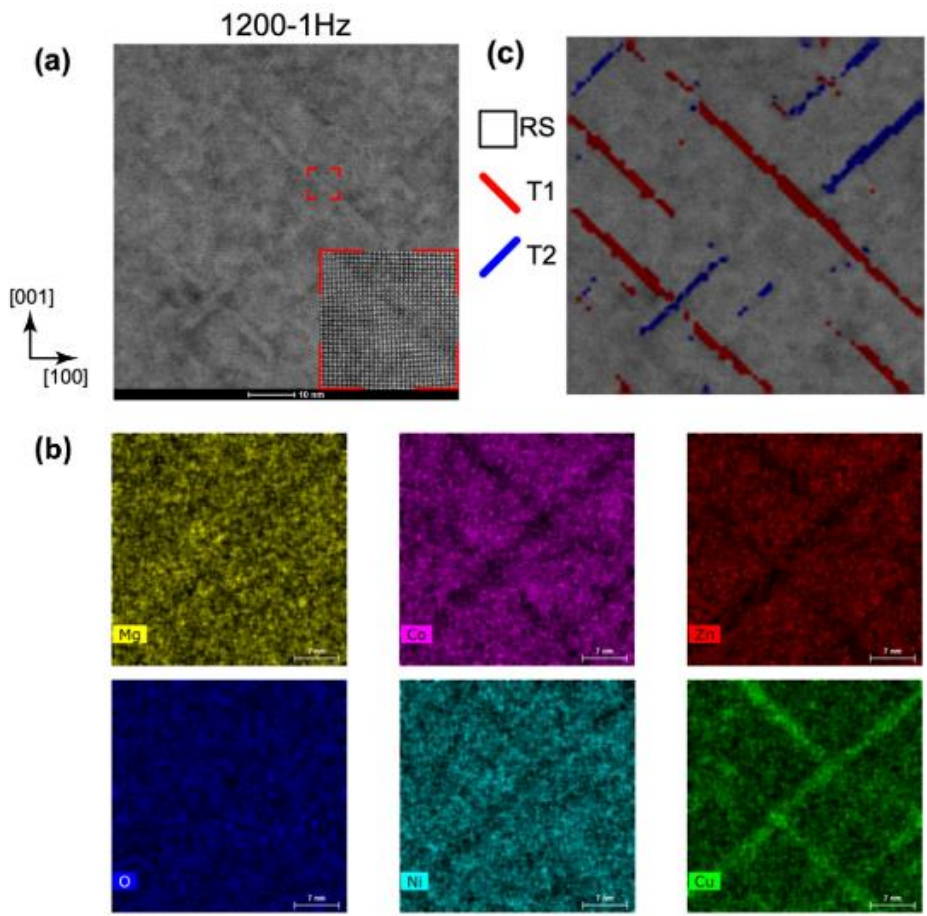

**Fig. 3:** Local structural and chemical analysis of $Mg_{0.2}Co_{0.2}Ni_{0.2}Cu_{0.2}Zn_{0.2}O$ grown at a 1 Hz pulse rate using 1200 total pulses: (a) ADF-STEM image and magnified inset of a tweed region; (b) representative EDX map; (c) results of image segregation based on local symmetry analysis with overlayed colors representing the rock salt matrix (RS, colorless) and tweed platelets oriented in two directions (T1 in red and T2 in blue)

**Local symmetry and chemical analysis: (b) Spinel-like nanocuboids**

Fig. 4a shows a higher-magnification atomically-resolved ADF-STEM image for the 6000-5Hz film highlighting one of the nano-cube regions. The magnified inset reveals additional periodic cation columns inside the cuboid. These locations correspond to the tetrahedral interstitials of the oxygen sublattice which in rock salt ($Fm\bar{3}m$ symmetry) are unoccupied. Populating these sites periodically will produce the spinel structure ($Fd\bar{3}m$ symmetry). There is precedent for such spinel-type clusters in rock salt matrices as in defect clusters in wüstite $Fe_{1-x}O$[40] and lithium manganese oxide[41]. Local symmetry analysis was applied to categorize the local Fourier transform patterns collected from Fig. 4a into two classes: RS, a rock salt matrix with no apparent nanostructuring; and SP, a spinel structured region (Fig.4 b) **(described in detail in Supplementary Note 2).** Additional analysis on the spinel nano-cuboids reveal a 2% lattice parameter reduction in both in- and out-of-plane directions not detected by XRD analysis.

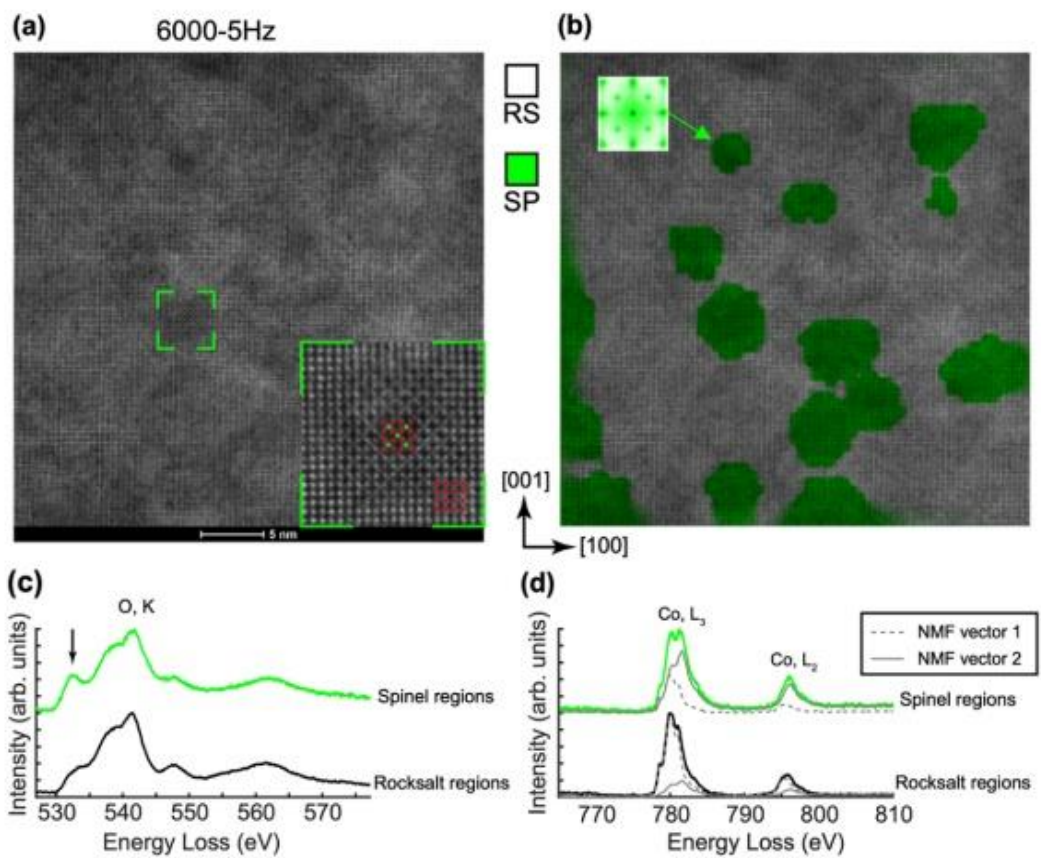

**Fig. 4:** Local structural and chemical analysis of $Mg_{0.2}Co_{0.2}Ni_{0.2}Cu_{0.2}Zn_{0.2}O$ grown at a 5 Hz pulse rate using 6000 total pulses: (a) ADF-STEM image and magnified inset of a spinel nano-cuboid region; (b) results of image segregation based on local symmetry analysis with overlayed colors representing rock salt matrix (RS, colorless) and spinel nano-cuboids (SP, green) with SP class averaged Fourier transform shown in the inset; (c) class-averaged O K edge ELNES; and (d) class-averaged Co L edge ELNES from the SP regions (green) and RS regions (black). Class-averaged ELNES spectra were decomposed as the linear combination of two NMF loading vectors, plotted in (d) with dashed and solid gray lines. These NMF vectors bear a striking resemblance to CoO and $Co_3O_4$ reference spectra (Supplementary note 3).

However, XRD shows two strong peaks in 2θ indicating epitaxial material with the expected pristine rock salt lattice constant and with a 0.5% smaller out-of-plane lattice constant. We attribute the second XRD peak to the heavily nanostructured rock salt matrix near the substrate interface **(Supplementary Note 1)**.

Ideal spinel structures exhibit $A_3O_4$ stoichiometry, necessitating some fraction of 3+ cations. Thus, we employed spatially resolved monochromated STEM-EELS to probe the energy loss near edge structure (ELNES) of the transition metal elements. Class-averaged EELS for the oxygen K edge ELNES and Co L edge ELNES spectra are plotted in Fig. 4c,d. An additional peak in the oxygen K spectra appears at approximately 531 eV in spinel-containing regions in Fig. 4c, indicating additional empty states in the transition metal 3d orbitals compared to RS regions. Ni and Cu ELNES spectra show no significant change **(Supplementary Note 3)**. However, the Co ELNES of SP regions closely resembles the reference ELNES spectra of spinel $Co_3O_4$, while the RS regions agrees well with the rock salt CoO ELNES reference[42,43]. These results suggest $Co^{3+}$ ions are responsible for accommodating the local symmetry change and producing the additional empty 3d states found in the oxygen K spectra.

If a slow growth rate and a large film thickness are critical ingredients for forming CuO nano-tweeds and spinel epitaxial nano-cuboids respectively, a thick and slowly grown film should possess both features. Fig. 5 depicts the TEM analysis of this 6000-1Hz sample. The SAED pattern (Fig. 5a) taken from a location near to the substrate interface shows streaking along <101> in reciprocal space and extra reflections at $\vec{g} = \{101\}$ validating our hypothesis. The $\vec{g}_{002}$ and $\vec{g}_{10\bar{1}}$ reflections were isolated to collect DF-TEM micrographs that reveal the spatial distribution of tweed platelets (Fig. 5b) and spinel nano-cuboids (Fig. 5c) respectively. Fig.5d shows the TEM image segregation results with both nanostructures. **(more analysis in Supplementary Note 1, 2 and 4)**.

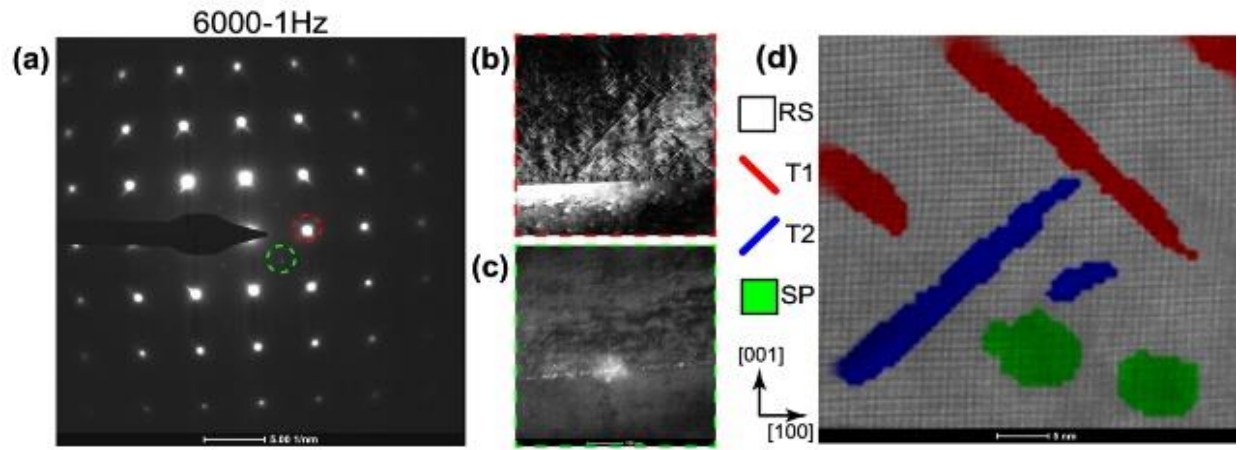

**Fig. 5:** Structural and chemical analysis of $Mg_{0.2}Co_{0.2}Ni_{0.2}Cu_{0.2}Zn_{0.2}O$ grown at a 1 Hz pulse rate using 6000 total pulses: (a) SAED pattern acquired at the interface of the film; (b) and (c) DF-TEM images showing the tweed structures and spinel nano-cuboids respectively, acquired using and reflections [red and green circles in (a)]; (d) results of image segregation based on local symmetry analysis with overlayed colors representing rock salt matrix (RS, colorless), tweed structures oriented in two directions (T1 in red and T2 in blue), and spinel nano-cuboids (SP, green).

**Multi-physics modeling for evolving order**

Nanostructure evolution is accompanied by the formation of interfaces, inhomogeneous stress/strain fields, and inhomogeneous chemical compositions, and it evolves to minimize the total free energy of the system comprising interfacial energy, strain energy, and chemical energy. To predict and analyze the morphological evolution, we developed a phase-field model for high-entropy oxides by incorporating the key thermodynamic contributions and DFT calculated lattice parameters and elastic constants. Temporal evolution and growth of the CuO-nanotweeds and spinel nano-cuboids from phase-field simulations are depicted in Fig. 6. The predicted morphological evolution predicts with experimental observations by STEM in Fig. 3 and 4 in both geometry and dimension. Strain mapping and total energy analysis for both nanostructures are presented and discussed in detail in **Supplementary Note 5**. Overall, phase-field simulations support the hypothesis that elastic strain energy and chemical potential reduction is a nanostructure influencer that drives exsolution in $Mg_{0.2}Co_{0.2}Ni_{0.2}Cu_{0.2}Zn_{0.2}O$.

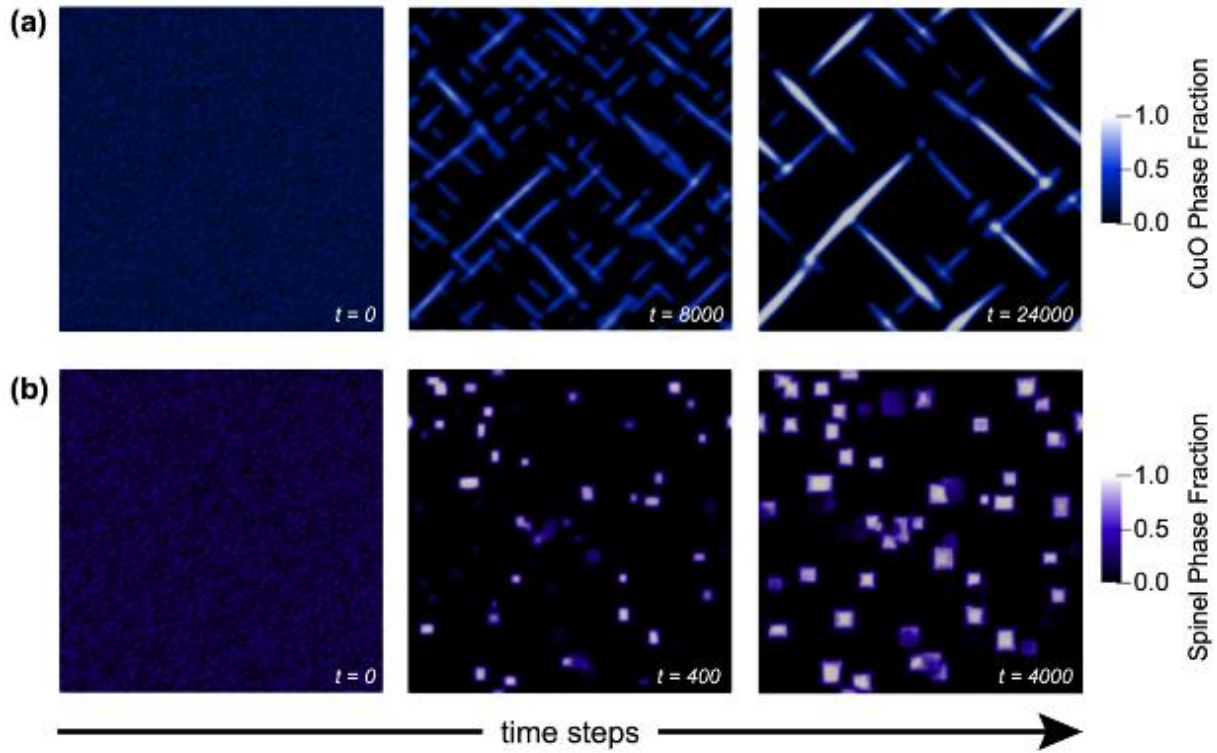

**Fig. 6:** Phase-field simulations of the morphologies of (a) CuO-rich tweed patterns, and (b) spinel nanocuboids in $Mg_{0.2}Co_{0.2}Ni_{0.2}Cu_{0.2}Zn_{0.2}O$. Cross-sectional images from 3D simulations are captured at y = 50nm. Temporal evolutions are carried out with discretized time steps of fixed lengths. Number of time steps for the displayed frames are labelled with inset text.

The CuO segregation and redistribution is driven by the chemical energy reduction overcoming the increase in the interfacial energy and the coherency strain energy associated with the coherent nanostructure. We only observe CuO in thin films where inter-pulse anneal is 5x longer. This suggests that kinetics are limited by the slow diffusion of CuO in the rock salt matrix. The CuO tweed shape (Fig. 6a) is a consequence of anisotropic stress-free transformation strain (SFTS) arising from the tetragonal Jahn-Teller distortion of the CuO relative to the cubic rock salt matrix. The simulated nano-tweed structure tilt is 43° relative to rock salt (001), in quantitative agreement with our TEM analysis (45°) and microelasticity theory calculations (44°)[22].

The spinel-nanocuboids are only observed in coherent thick films where the elastic strain energy per unit area is ~5 times larger than in the thinner films. The large compressive elastic strain energy in the thicker films favors $Co^{3+}$ due to its ~18% smaller ionic radii compared to $Co^{2+}$.[43] A complementary explanation is that Co favors $3^+$ state at our growth conditions (400 °C, but well below the solid solution stability threshold of 875 °C and a high oxygen chemical potential)[43] thus the increased strain within thick films mediates its transition from quenched and trapped $Co^{2+}$ to enthalpy-favored $Co^{3+}$. We speculate that the following defect chemical process accompanies this conversion to conserve charge, mass, and lattice site ratios:

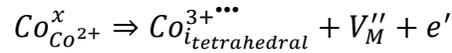

$$Co_{Co^{2+}}^{x} \Rightarrow Co_{i_{tetrahedral}}^{3+\cdots} + V_M'' + e'$$

This site occupancy and oxidation state change of Co into the newly unlocked tetrahedral sites produces the chemically-driven and enthalpy-favored spinel phase. Collectively, the inclusion of $Co^{3+}$-containing spinel nanostructure within the rock salt matrix affords an overall reduction of strain energy for the thicker films **(discussed in detail in Supplementary Note 5)**. The cubic-to-cubic transformation of spinel nano-cuboids induces isotropic SFTS, resulting in overall smaller elastic anisotropy compared to the anisotropic SFTS of CuO-tweeds (cubic-to-tetragonal). The elastic anisotropy of the spinel phase is due to the anisotropic elastic stiffness, which is defined for a cubic crystal as[22]:

$$\zeta = \frac{c_{11}-c_{12}-2c_{44}}{c_{44}}.$$

$\zeta$ = -0.055 for rock salt $Mg_{0.2}Co_{0.2}Ni_{0.2}Cu_{0.2}Zn_{0.2}O$, while $\zeta$ = -1.117 for the $Co_3O_4$ spinel. An additional major anisotropic contribution comes from the estimated low-index interfacial energies ({100}: 0.92 J/m², {110}: 1.31 J/m², {111}: 2.31 J/m²). When a crystal phase evolves in a constrained matrix, as in the present topotactic transformation, a more negative $\zeta$ parameter and strong anisotropic interfacial energies both favor the habit planes with {100} low index facets, explaining the cuboidal precipitate morphology in Fig.2k and Fig.6b.

## Discussion and Outlook

The slow-to-fast and thin-to-thick deposition conditions explored in this PLD study provide kinetic access to an interesting set of topotactic nanostructures that evolve during experimental timescales. They occur through a chemical ordering process where either Cu or Co chemically segregate into lattice sites that offer a lower overall free energy at the expense of entropic disorder.

While phase separation from a metastable solid solution to a multiphase ensemble is both expected and well known, exploring such processes in an initially entropy-stabilized host offers new insights into material engineering. $Mg_{0.2}Co_{0.2}Ni_{0.2}Cu_{0.2}Zn_{0.2}O$ illustrates a diversity of chemical, structural, and defect chemistry nanostructures that involve both continuous (*i.e.*, ordering while preserving coherent boundaries) and discontinuous formation paths. Prior work shows how thin film synthesis temperature alone can produce a 5% lattice constant swing without a structure transition while the present work details two distinct nanostructure ordering motifs. The interesting difference separating the possibilities is the degree and type of energy available to initiate order, and the thermodynamic boundary conditions (*i.e.*, oxygen activity and strain) that can bias the transition along a preferred path. It is difficult to find crystalline systems with simpler chemistry that exhibit such behavior. While non-existence cannot prove a hypothesis, in this case it supports the possibility that high entropy crystals, by virtue of their enormous configuration space, can access myriad microstates with a continuum of closely spaced free energies (as they serially adopt small degrees of order) that create pathways to the nanostructured crystals presently observed. It is additionally interesting to note the macroscopic tolerance for local disorder in these crystals. X-ray diffraction (Fig. 2a) shows examples where $Mg_{0.2}Co_{0.2}Ni_{0.2}Cu_{0.2}Zn_{0.2}O$ film is fully populated by spinel pockets with ~1% misfit strain, but x-ray line widths remain as narrow as the homogeneous parent solid solution, and the film remains pseudomorphic with almost no misfit interfacial dislocations. It appears possible that the diversity of local ordered and disordered structures provides different mechanisms to minimize energy, which may be the unique, scientifically interesting, and technology enabling cornerstone of the entropy-stabilized formulation. Furthermore, it is also interesting and comforting that existing principles of crystal chemistry and thermodynamics can predictively explain this entire observation set. Phase-field models capture the exsolved nanostructures with high chemical and structural fidelity.

While the entropy-stabilized random and homogeneous solid solution is the highest-entropy member of the structural continuum and it receives most experimental attention, it represents only one macrostate. In this manuscript we show that in $Mg_{0.2}Co_{0.2}Ni_{0.2}Cu_{0.2}Zn_{0.2}O$, and likely many other high entropy formulations, it is possible to deliberately and predictively "spend" entropy by synthesis, thermal history and boundary condition combinations to metastabilize different macrostates along the ordering spectrum. We propose that each one represents a new material with a likelihood of new property opportunities, and that the key to active material engineering in high-configuration space is understanding how formulation, structure, and process combine to regulate entropic spending.

## Methods

*Sample Synthesis*

$Mg_{0.2}Co_{0.2}Ni_{0.2}Cu_{0.2}Zn_{0.2}O$ epitaxial films were grown using similar methods to those reported previously[38]. A single-phase ceramic ablation target was reactively sintered from a stoichiometric binary oxide powder mixture MgO (≥99%), CoO (≥99%), NiO (99.99%), CuO (99.99%), and ZnO (99.999%) in air at 1000 °C for 12 hours and air-quenched to obtain a single rock salt phase. Thin films were deposited on [001]-MgO substrates by PLD using a Coherent 248 nm KrF laser. Substrates were cleaned with acetone, isopropanol, and methanol followed by a 10-minute UV-ozone treatment then were adhered to a block heater with silver paint (Ted Pella Leitsilber 200). The substrate temperature was 400 °C to ensure relatively low $T_{sub}$ ~400 °C to preserve metastability and to avoid low-temperature enhanced $Co^{3+}$ concentrations[43]. 25 sccm oxygen $O_2$ was flowed into the chamber, the gate valve was throttled until the chamber pressure stabilized at 50 mTorr. Laser fluence was 1.2 J/cm$^2$ with a total laser pulse energy of 170 mJ. All films were quenched to ambient lab conditions by venting the chamber to ambient air less than 2 minutes after the final laser pulse to mitigate post-synthesis annealing effects.

*Sample Characterization*

X-ray diffraction (XRD) was performed using a PANalytical Empyrean diffractometer. The primary incident beam optic was a hybrid mirror/2-bounce Ge monochromator combination, the primary diffracted beam optic was a programmable divergent slit and a PIXcel$^{3D}$ detector combination. Transmission electron microscopy (TEM) samples were fabricated using a focused ion beam (FIB) approach. First, a 500 nm carbon layer was deposited with the electron beam to protect the top surface. A second 4 µm layer was subsequently deposited with the ion beam. Thin TEM lamella were milled using a 30 kV Ga ion beam and lifted out *in situ* and mounted on a carbon FIB grid to avoid contamination. The samples were thinned with 30 kV and 5 kV Ga ion beams to electron transparency. Resulting lamellae were then cleaned with 2 kV and 1 kV Ga ion beams to reduce damage and amorphization.

Selected area electron diffraction (SAED) patterns and dark field TEM (DF-TEM) images were collected using a ThermoFisher Talos 200X TEM at 200 kV. Annular dark-field scanning transmission electron microscopy (ADF-STEM) images, low angle annular dark-field STEM (LAADF-STEM) images, energy dispersive X-ray (EDX) spectra, and electron energy loss spectra (EELS) were collected using a ThermoFisher Titan G2 aberration-corrected STEM at 300 kV. The convergence angle was 21.5 mrad, and the collection angle was 61-200 mrad for ADF-STEM mode and 26-149 mrad for LAADF-STEM mode. EELS spectra were collected in the dual-EELS mode with an energy resolution around 0.18 eV, measured with the full-width half maxima of the zero-loss peak (ZLP). The EELS collection angle was 12.2 mrad.

TEM image analysis employed a physics-based deep data image analysis algorithm that can categorize local symmetry from atomically resolved imaging data to extract *a priori* unknown local distortions[44,45]. Subsequent data classification employed dimensionality reduction via principal component analysis (PCA), followed by data clustering using a gaussian mixture model (GMM). A flowchart illustrating our procedure is shown in Fig. S1. A cropping window with a height (h) and width (w) of 500 pixels was placed on an ADF-STEM image. The center position of the window was shifted across the entire image with the step size of 12 pixels, and sub-images were cropped at each position forming a stack of sub-images with three dimensions (h by w by n). The positions of the cropping window were recorded as (xp, yp). Next, a 2D Fourier transformation (FT) was performed on each sub-image, and this step generated a stack of FT patterns that represent the local periodicity of the crystal. The stack of FT patterns was then flattened into two dimensions (h • w by n). For PCA, the stack of FT patterns was categorized into four, three or two classes for STEM images collected from 6000-1Hz, 1200-1Hz or 6000-5Hz films, respectively. Class-averaged FT patterns were extracted by averaging the FT patterns within each class. Finally, image segregation was performed by projecting the corresponding cropping window positions back to the original STEM image and overlaid with color based on the class.

The STEM-EELS analysis was performed using an unsupervised machine learning algorithm. Each EELS spectrum was corrected based on the ZLP position. The unsupervised machine learning algorithm included dimensionality reduction with nonnegative matrix factorization (NMF) and clustering with GMM using the *fitgmdist* MATLAB function[46]. During the STEM-EELS experiment, three-dimensional spectral image data (h by w by 2048 channels) was acquired from the area of interest. The spectral image was flattened into two dimensions (h•w by 2048), and the dimensionality reduction was performed to decompose the data into a linear combination defined by two loading vectors (2 by 2048) and the corresponding scores (h•w by 2). The clustering with GMM was performed on the NMF scores, and the spectral image data was clustered into two classes. The class-averaged spectra were then extracted based on the clustering results.

*Computational Methods*

The phase-field method is a mesoscale density-based computational method that allows modeling microstructural temporal evolution during materials processes **(Supplementary Note 6)**.[47] Phase-field simulations were carried out with the semi-implicit Fourier-spectral method[48] which has been applied to solve the time-dependent Allen-Cahn and diffusion equations with periodic boundary conditions as described in the supplemental materials[22,47]. We note that our models capture the growth and coarsening stages of two nanostructure types within the rock salt matrix. Each simulation minimizes the total free-energy as a function of order parameters representing the phase distributions and cation concentrations. The order parameters initial distribution is given by random fields, while the initial chemical concentrations are homogeneous for each simulation. 3D simulations of (120 nm)$^3$ were used to model the CuO-rich nano-tweed and spinel nano-cuboids, which are discretized into 1 nm$^3$ cubic grids. A 2D slice was utilized for the nano-tweed for ease of interpretation. The total free-energy considers the chemical potentials, elastic strain energies, and interfacial energies. Elastic stiffness and stress-free strain values were calculated using density functional theory (DFT) **(Supplementary Note 7)**. For the CuO-rich tweed simulations, we employed experimental strain values due to overestimation of stress-free DFT calculations. The magnitude of rock salt-spinel interfacial energies are approximated using the reported value for {100} CoO / {100} Co$_3$O$_4$ from Chen *et al.*[49], while the anisotropy is incorporated using relative ratios of Co$_3$O$_4$ low-index surface energies[50].

# References

1.	Ranganathan, S. Alloyed pleasures: Multimetallic cocktails. *Curr. Sci.* **85**, 1404–1406 (2003).

2.	Murty, B. S., Yeh, J.-W., Ranganathan, S. & Bhattacharjee, P. P. *High-Entropy Alloys*. (Elsevier, 2019).

3.	Kotsonis, G. N. *et al.* High-entropy oxides: Harnessing crystalline disorder for emergent functionality. *Journal of the American Ceramic Society* **106**, 5587–5611 (2023).

4.	Hernandez, F. C. R., Ramírez, J. M. H. & Mackay, R. *Al-Si alloys: automotive, aeronautical, and aerospace applications*. (Springer, 2017).

5.	Singh, R., Oprysko, M. M. & Harame, D. Silicon germanium: technology, modeling, and design. (2004).

6.	Farrell, R. M. *et al.* Low-threshold-current-density AlGaN-cladding-free m-plane InGaN/GaN laser diodes. *Applied Physics Letters* **96**, 231113 (2010).

7.	Li, F. *et al.* Giant piezoelectricity of Sm-doped Pb(Mg1/3Nb2/3)O3-PbTiO3 single crystals. *Science* **364**, 264–268 (2019).

8.	Fichtner, S., Wolff, N., Lofink, F., Kienle, L. & Wagner, B. AlScN: A III-V semiconductor based ferroelectric. *Journal of Applied Physics* **125**, 114103 (2019).

9.	Hayden, J. *et al.* Ferroelectricity in boron-substituted aluminum nitride thin films. *Phys. Rev. Mater.* **5**, 044412 (2021).

10.	Adachi, S. GaAs, AlAs, and AlxGa1−xAs: Material parameters for use in research and device applications. *Journal of Applied Physics* **58**, R1–R29 (1985).

11.	Berlincourt, D., Cmolik, C. & Jaffe, H. Piezoelectric Properties of Polycrystalline Lead Titanate Zirconate Compositions. *Proc. IRE* **48**, 220–229 (1960).

12.	Goldschmidt, V. The laws of crystal chemistry. *Naturwissenschaften* **14**, 477–485 (1926).

13.	Hume-Rothery, W. & Powell, H. M. On the theory of super-lattice structures in alloys. *Zeitschrift für Kristallographie-Crystalline Materials* **91**, 23–47 (1935).

14.	Pauling, L. The sizes of ions and the structure of ionic crystals. *Journal of the American Chemical Society* **49**, 765–790 (1927).

15.	Megaw, H. D. *Crystal Structures: A Working Approach*. (TechBooks, 1973).

16.	Muller, O. & Roy, R. *The Major Ternary Structural Families*. (Springer-Verlag, 1974).

17.	Cantor, B., Chang, I. T. H., Knight, P. & Vincent, A. J. B. Microstructural development in equiatomic multicomponent alloys. *Materials Science and Engineering: A* **375–377**, 213–218 (2004).

18.	Yeh, J. *et al.* Nanostructured high-entropy alloys with multiple principal elements: novel alloy design concepts and outcomes. *Advanced engineering materials* **6**, 299–303 (2004).

19.	Rost, C. M. *et al.* Entropy-stabilized oxides. *Nature Communications* **6**, 8485 (2015).

20. Sarkar, A. *et al.* High-Entropy Oxides: Fundamental Aspects and Electrochemical Properties. *Adv Mater* **31**, e1806236 (2019).

21. Aamlid, S. S., Oudah, M., Rottler, J. & Hallas, A. M. Understanding the Role of Entropy in High Entropy Oxides. *J. Am. Chem. Soc.* **145**, 5991–6006 (2023).

22. Khachaturyan, A. Theory of Structural Transformations in Solids (Mineola, NY. (2008).

23. Rost, C. M., Rak, Z., Brenner, D. W. & Maria, J.-P. Local structure of the $Mg_xNi_xCo_xCu_xZn_xO(x=0.2)$ entropy-stabilized oxide: An EXAFS study. *Journal of the American Ceramic Society* **100**, 2732–2738 (2017).

24. Diercks, D. R., Brennecka, G., Gorman, B. P., Rost, C. M. & Maria, J.-P. Nanoscale Compositional Analysis of a Thermally Processed Entropy-Stabilized Oxide via Correlative TEM and APT. *Microscopy and Microanalysis* **23**, 1640–1641 (2017).

25. Bhaskar, L. K., Nallathambi, V. & Kumar, R. Critical role of cationic local stresses on the stabilization of entropy-stabilized transition metal oxides. *Journal of the American Ceramic Society* **103**, 3416–3424 (2020).

26. Dupuy, A. D., Wang, X. & Schoenung, J. M. Entropic phase transformation in nanocrystalline high entropy oxides. *Materials Research Letters* **7**, 60–67 (2019).

27. Zhang, J. *et al.* Long-Range Antiferromagnetic Order in a Rocksalt High Entropy Oxide. *Chem. Mater.* **31**, 3705–3711 (2019).

28. Pitike, K. C., KC, S., Eisenbach, M., Bridges, C. A. & Cooper, V. R. Predicting the Phase Stability of Multicomponent High-Entropy Compounds. *Chem. Mater.* **32**, 7507–7515 (2020).

29. Anand, G., Wynn, A. P., Handley, C. M. & Freeman, C. L. Phase stability and distortion in high-entropy oxides. *Acta Materialia* **146**, 119–125 (2018).

30. Musicó, B. *et al.* Tunable magnetic ordering through cation selection in entropic spinel oxides. *Phys. Rev. Mater.* **3**, 104416 (2019).

31. Witte, R. *et al.* High-entropy oxides: An emerging prospect for magnetic rare-earth transition metal perovskites. *Phys. Rev. Mater.* **3**, 034406 (2019).

32. Mazza, A. R. *et al.* Designing magnetism in high entropy oxides. *Advanced Science* **9**, 2200391 (2022).

33. Lun, Z. *et al.* Cation-disordered rocksalt-type high-entropy cathodes for Li-ion batteries. *Nature Materials* **20**, 214–221 (2021).

34. Zhai, S. *et al.* The use of poly-cation oxides to lower the temperature of two-step thermochemical water splitting. *Energy & Environmental Science* **11**, 2172–2178 (2018).

35. Kotsonis, G. N., Rost, C. M., Harris, D. T. & Maria, J.-P. Epitaxial entropy-stabilized oxides: growth of chemically diverse phases via kinetic bombardment. *MRS Communications* **8**, 1371–1377 (2018).

36. DAVIES, P. K. Investigations of NiO-CuO Solid Solutions Using Transmission Electron Microscopy: I, Tweed Microstructure. *Journal of the American Ceramic Society* **69**, 796–799 (1986).

37.	Zhu, Y. & Cai, Z.-X. Diffuse scattering and image contrast of tweed in superconducting oxides: simulation and interpretation. *Ultramicroscopy* **52**, 539–548 (1993).

38.	Berardan, D., Meena, A. K., Franger, S., Herrero, C. & Dragoe, N. Controlled Jahn-Teller distortion in (MgCoNiCuZn)O-based high entropy oxides. *Journal of Alloys and Compounds* **704**, 693–700 (2017).

39.	Rák, Zs., Maria, J.-P. & Brenner, D. W. Evidence for Jahn-Teller compression in the (Mg, Co, Ni, Cu, Zn)O entropy-stabilized oxide: A DFT study. *Materials Letters* **217**, 300–303 (2018).

40.	Hazen, R. M. & Jeanloz, R. Wüstite ($Fe_{1-x}$ O): A review of its defect structure and physical properties. *Reviews of Geophysics* **22**, 37–46 (1984).

41.	Gao, P. *et al.* Atomic-Scale Tracking of a Phase Transition from Spinel to Rocksalt in Lithium Manganese Oxide. *Chem. Mater.* **29**, 1006–1013 (2017).

42.	Meisenheimer, P. B. *et al.* Magnetic frustration control through tunable stereochemically driven disorder in entropy-stabilized oxides. *Phys. Rev. Mater.* **3**, 104420 (2019).

43.	Kotsonis, G. N. *et al.* Property and cation valence engineering in entropy-stabilized oxide thin films. *Phys. Rev. Mater.* **4**, 100401 (2020).

44.	Vasudevan, R. K., Ziatdinov, M., Jesse, S. & Kalinin, S. V. Phases and Interfaces from Real Space Atomically Resolved Data: Physics-Based Deep Data Image Analysis. *Nano Lett.* **16**, 5574–5581 (2016).

45.	Somnath, S. *et al.* Feature extraction via similarity search: application to atom finding and denoising in electron and scanning probe microscopy imaging. *Advanced Structural and Chemical Imaging* **4**, 3 (2018).

46.	McLachlan, G. J., Lee, S. X. & Rathnayake, S. I. Finite Mixture Models. (2019).

47.	Chen, L.-Q. & Zhao, Y. From classical thermodynamics to phase-field method. *Progress in Materials Science* **124**, 100868 (2022).

48.	Chen, L. Q. & Shen, J. Applications of semi-implicit Fourier-spectral method to phase field equations. *Computer Physics Communications* **108**, 147–158 (1998).

49.	Chen, X., Van Gog, H. & Van Huis, M. A. Transformation of $Co_3O_4$ nanoparticles to CoO monitored by in situ TEM and predicted ferromagnetism at the $Co_3O_4$/CoO interface from first principles. *Journal of Materials Chemistry C* **9**, 5662–5675 (2021).

50.	Ribeiro, R. A. P., De Lazaro, S. R., Gracia, L., Longo, E. & Andrés, J. Theoretical approach for determining the relation between the morphology and surface magnetism of $Co_3O_4$. *Journal of Magnetism and Magnetic Materials* **453**, 262–267 (2018).

# Acknowledgments

The authors gratefully acknowledge support from NSF MRSEC DMR-2011839. The authors would also like to acknowledge many helpful and enlightening discussions with Susan Trolier-McKinstry, John T. Heron, Venkatraman Gopalan, and Zhiqiang Mao regarding the nuances of high-entropy systems. Phase-field simulations and DFT calculations utilized resources from the ROAR cluster of the Penn State Institute for Computational and Data Sciences.

# SUPPLEMENTAL INFORMATION

**Supplementary Note 1: Nanocuboids distribution**

The 6000-5Hz film LAADF-STEM mode shown in Fig.2k reveals a high density of nanoscopic cuboids, indicating local lattice strain at boundaries between cuboids and the matrix. The cuboids appear to be distributed randomly throughout the rock salt matrix. However, a closer examination reveals a slightly higher density near the film-substrate interface (Fig. S1.a). The same can be observed for 6000-1Hz film (Fig. S1.b). It is worth noting that there appears a lower concentration of nanocuboids in the 6000-1Hz film which could be due to the presence of the second exsolution mechanism of CuO nanotweeds. In fact, the nano-tweed structures in 6000-1Hz film are present through the film thickness besides a thin layer very close to the substrate while the spinel structures are mostly located within 20 nm of the substrate.

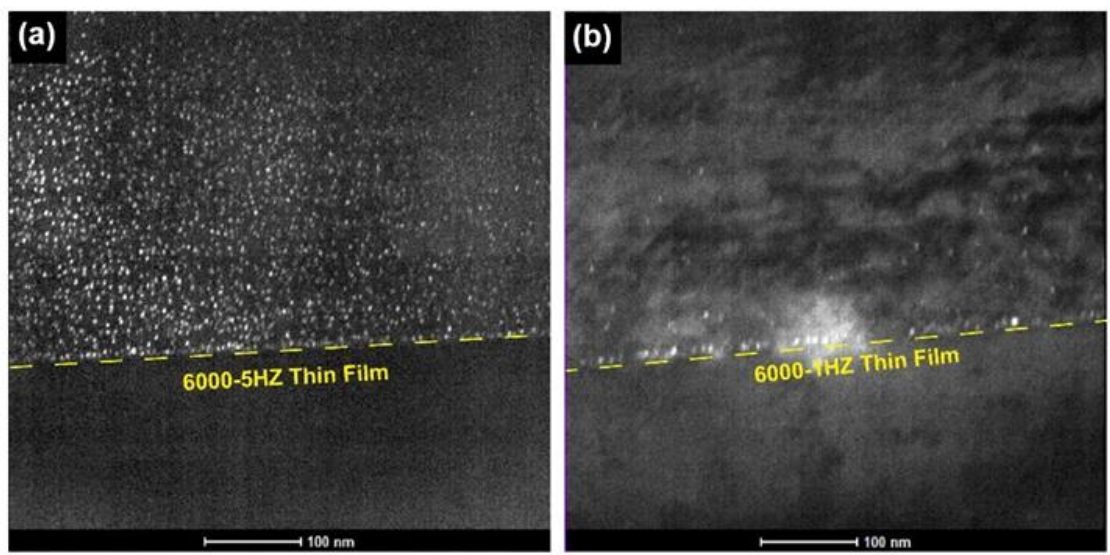

Fig. S1: DF-TEM images showing spinel nano-cuboids in both thick films. In (b) the nano tweeds can be observed as well.

**Supplementary Note 2: Local symmetry analysis**

Fig. S2 illustrates the machine learning algorithm we applied for the image segregation. Fig. S3.a-c shows the actual class-averaged FT patterns and the differences between them for the 1200-1Hz, 6000-5Hz and 6000-1Hz films, respectively. For the 1200-1Hz film, T1 and T2 regions both exhibit a contraction of $\vec{g} = 002$ and extension of $\vec{g} = 200$ relative to the RS matrix, indicating a local elongation of the out-of-plane lattice parameter by around 5% and shrinkage of the in-plane lattice parameter by around -3%; consistent with observations of CuO Jahn-Teller distortions. The differences between class-averaged FT patterns (FigS3.a bottom left) suggest a lattice rotation of roughly 5° in opposite directions for T1 and T2 which are not evident in the XRD pattern (Fig. 2a and Fig. 2c), signifying the globally averaged lattice parameter is dominated by the rock salt matrix. Strain mapping is illustrated and discussed in **Supplementary Note 5**. The 6000-5Hz film (Fig. S3b) exhibits extra diffraction spots at $\vec{g}$ = 101, 10$\bar{1}$, $\bar{1}$01, and $\bar{1}$0$\bar{1}$, the same set of extra spots observed by SAED (Fig. 2h). These spots underline the local symmetry change from rock salt to spinel. A slight extension of the $\vec{g}$ = 002 and $\vec{g}$ = 200 is visible for the spinel nano-cuboids, indicating a 2% lattice parameter reduction in both in- and out-of-plane directions (**Supplementary Note 5**). XRD scans do not capture this reduction, which is normally easily detected, suggesting that the distortions exist over very short ranges. Fig. S3c shows the RS-T1, RS-T2 and RS-SP for the 6000-1Hz film which indicates distortions similar to those observed in both the 1200-1Hz and 6000-5Hz samples.

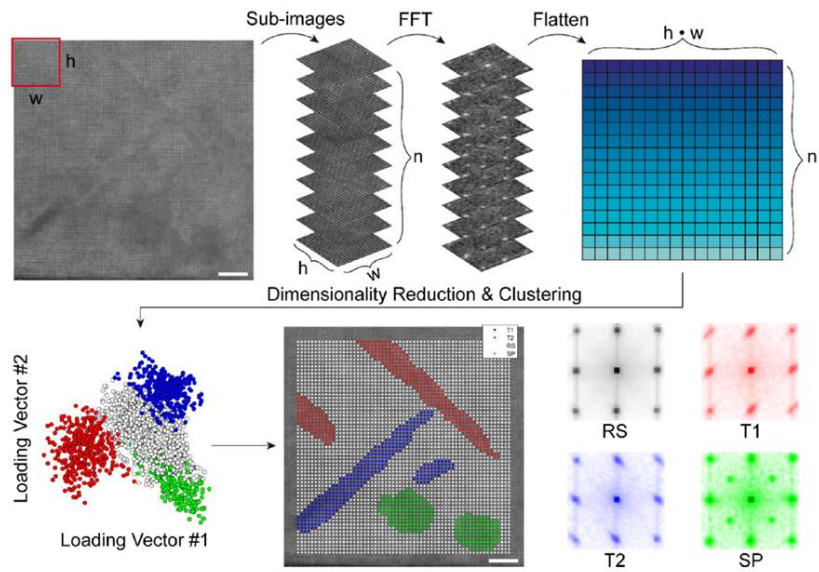

Fig. S2: Flowchart illustrating the unsupervised machine learning algorithm employed for image segregation.

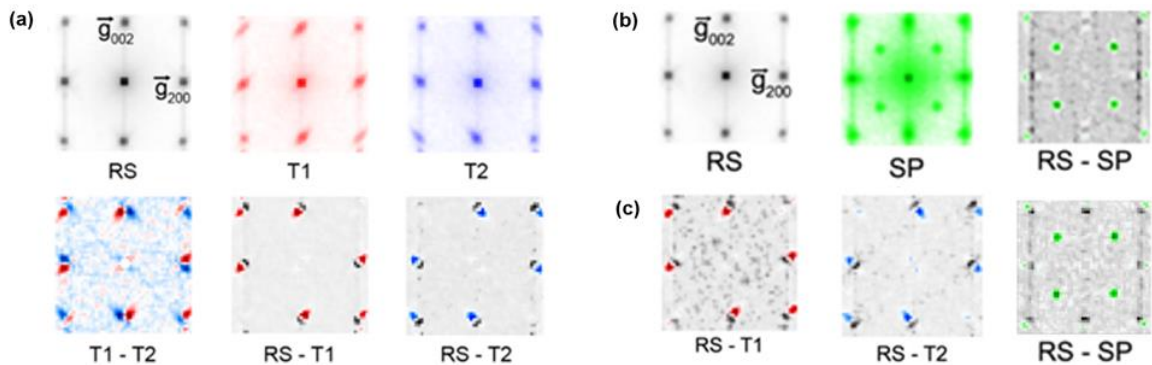

Fig. S3: Class-averaged FT patterns for each class; and differences between the class-averaged FT patterns highlighting the local lattice distortions for the 1200-1Hz (a), 6000-5Hz (b) and 6000-1Hz (c) films.

## Supplementary Note 3: EELS analysis for the 6000-5Hz film

Fig. S4 shows the data analysis on EELS O K edge. Fig. S5 shows the data analysis on EELS Co L edge. The two NMF loading vectors bear a striking resemblance to CoO and $Co_3O_4$ reference spectra and previous X-ray absorption results for $Mg_{0.2}Co_{0.2}Ni_{0.2}Cu_{0.2}Zn_{0.2}O$ with varying $Co^{2+}/Co^{3+}$ ratios[1,2]. Fig. S6 shows that Ni and Cu spectra have no noteworthy changes.

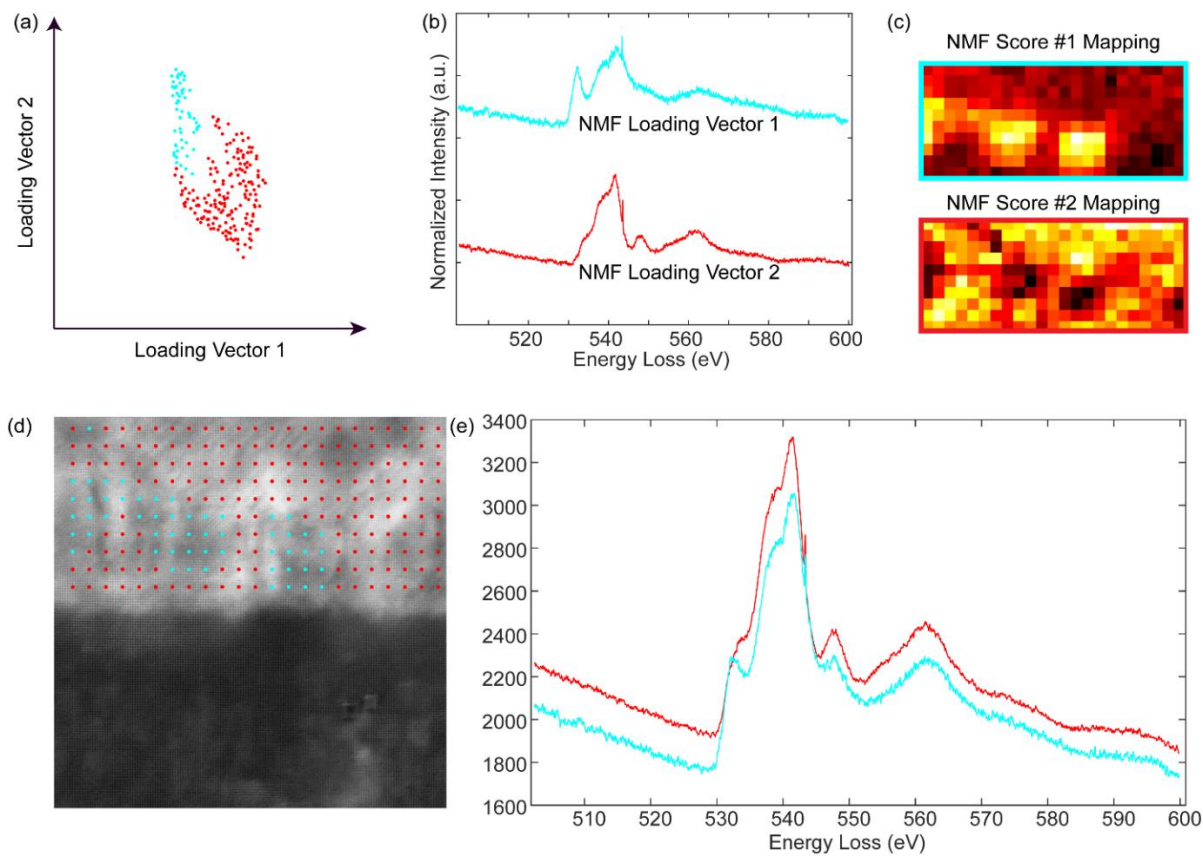

Fig. S4: Data analysis performed on EELS O K edge data acquired from rock salt and spinel nano-cuboid regions of the 6000-5Hz film[20] . (a) scatter plot of the coefficient of the two NMF loading vectors. The spectral data is clustered into two classes based on the coefficient; (b) plot of the two NMF loading vectors; (c) the heatmap showing the score of the two NMF loading vectors; (d) the probe positions overlaid on the simultaneously acquired ADF-STEM image; (e) the class-averaged O K edge.

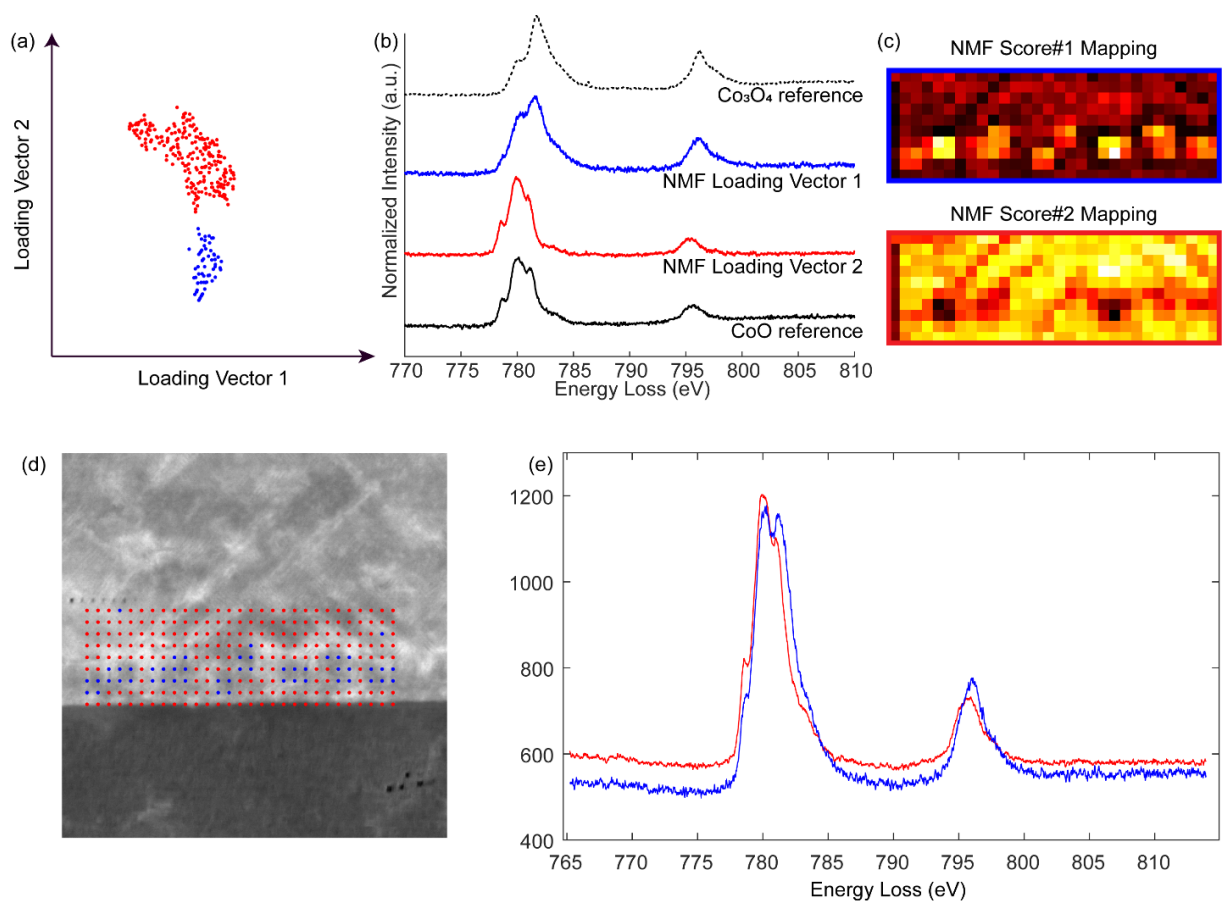

Fig. S5: Data analysis performed on EELS Co L edge data acquired from rock salt and spinel nano-cuboids regions of the 6000-5Hz film. (a) the scatter plot of the coefficient of the two NMF loading vectors. The spectra data is clustered into two classes based on the coefficient: (b) plot of the two NMF loading vectors and the reference ELNES for CoO and $Co_3O_4$. The NMF loading vectors show striking similarity compared to the reference ELNES; (c) the heatmap showing the score of the two NMF loading vectors; (d) the probe positions overlaid on the simultaneously acquired ADF-STEM image; (e) the class-averaged Co L edge.

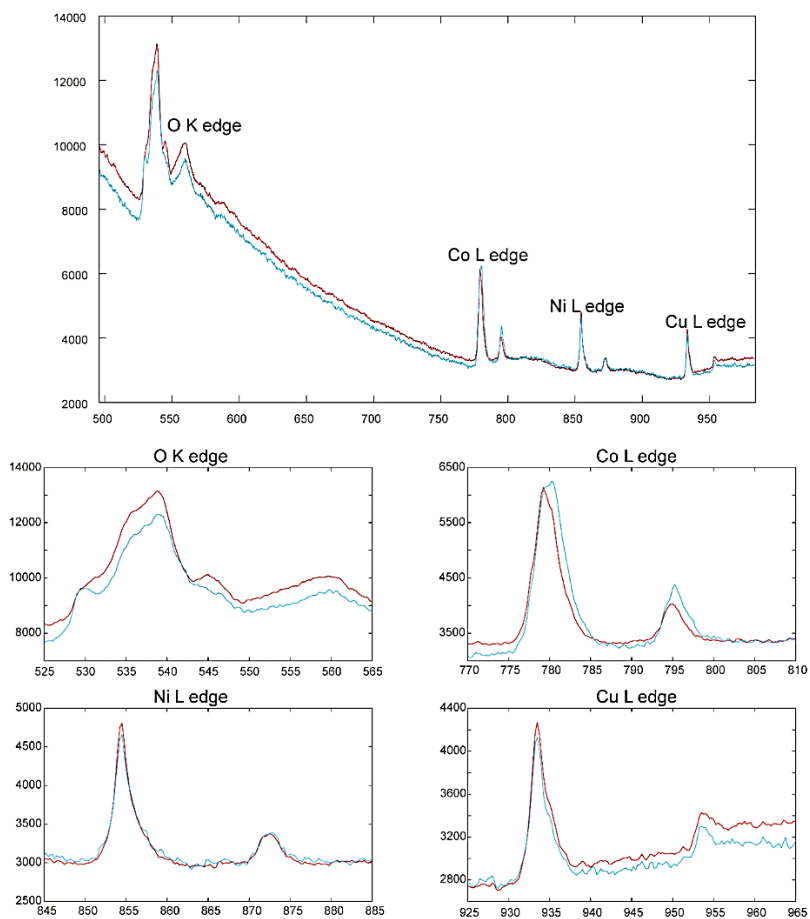

Fig. S6: EELS spectra acquired from rock salt (red) and spinel nano-cuboids (cyan) regions containing O K edge, Co L edge, Ni L edge and Cu L edge. In spinel nano-cuboids data, the O K edge spectrum shows an extra peak at 531 eV and Co L edge spectrum shows shift and intensity ratio variation. On the other hand, Ni and Cu L edge do not show significant changes.

## Supplementary Note 4: 6000-1Hz film EDX mapping

EDX mapping (Fig. S7) reveals the expected surplus of Cu ions in the tweed structures in the 6000-1Hz film.

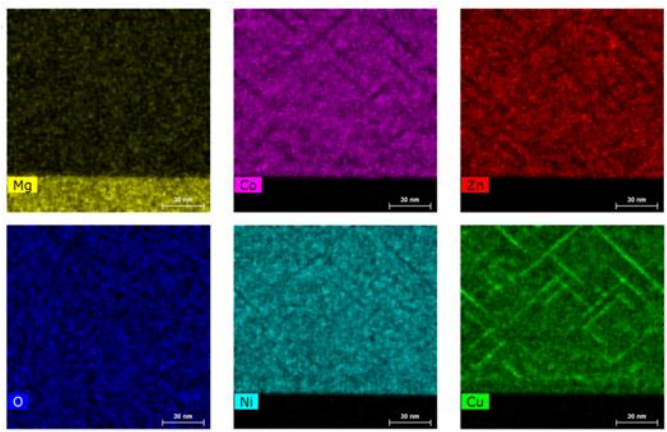

Fig. S7: EDX map of the 6000-1Hz film

**Supplementary Note 5: Nanostructures strain mapping and energy analysis**

Fig. S6 illustrates the TEM strain mapping for CuO-tweeds and the spinel-nanocuboids in the $Mg_{0.2}Co_{0.2}Ni_{0.2}Cu_{0.2}Zn_{0.2}O$ rock salt phase, including in-plane ($\varepsilon_{xx}$), out-of-plane ($\varepsilon_{zz}$), and rotational ($\omega_{xz}$) components, respectively. Fig. S7-9 show the remarkably reproduced and predicted strain maps from our phase-field modeling. Strain values are based on distortions from the $Mg_{0.2}Co_{0.2}Ni_{0.2}Cu_{0.2}Zn_{0.2}O$ pristine rock salt matrix. The depictions of the total energy minimization and the evolution of its components are shown in Fig. S10-12. In the CuO-tweed regions, strain values are substantially larger than those in the rock salt matrix, in agreement with TEM measurements (Fig. S6). Although the chemical energy is higher within the CuO-tweed (Fig. S11a), it is compensated by the elastic strain energy reduction in the matrix (Fig. S12a). For the spinel case, we observe a slightly different phenomenon: local strain redistribution via alteration of crystal symmetry and chemical environment. Phase-field simulations suggest that negative strain in the $Co^{3+}$-containing spinel cubes locally alleviates the matrix epitaxial strain imposed by the substrate, reducing the matrix elastic strain energy by chemical reconfiguration as shown in Fig. S12b.

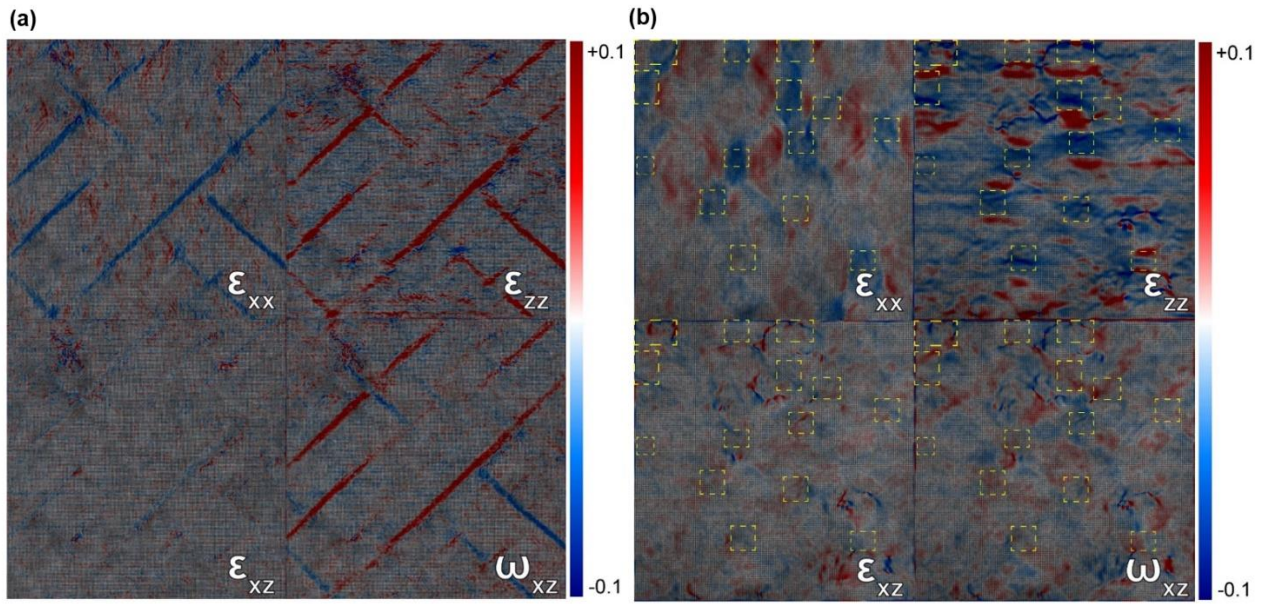

Fig. S6: Strain mapping with geometric phase analysis (GPA) on atomically resolved STEM images for (a) tweed nanostructures and (b) spinel nano-cuboids (yellow boxes).

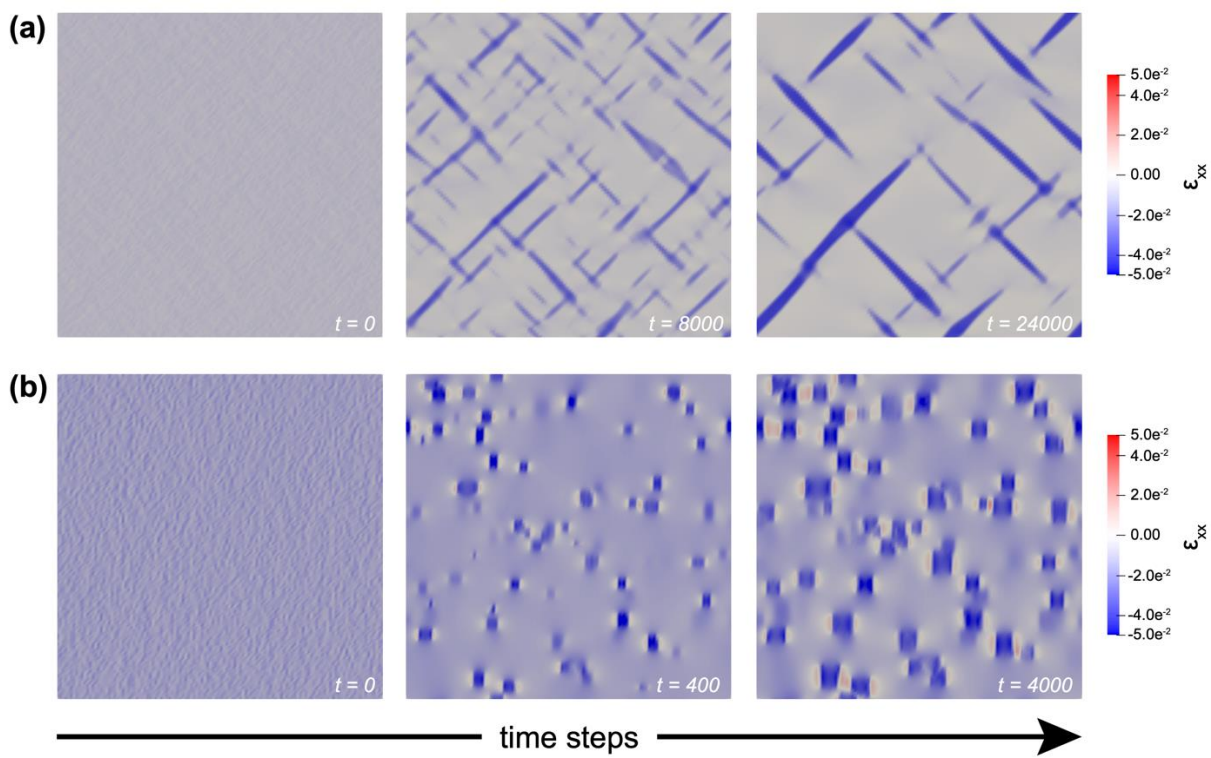

Fig. S7: Phase-field in-plane strain maps for (a) tweed nanostructures and (b) spinel nano-cuboids

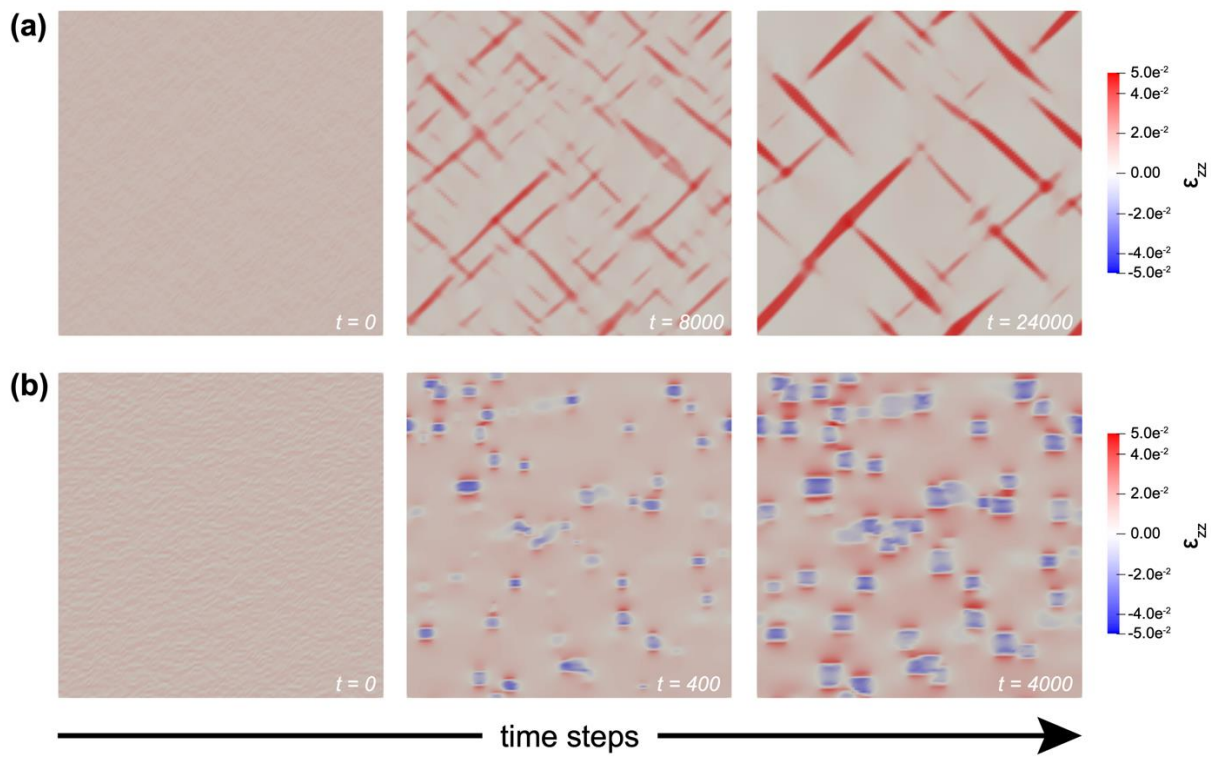

Fig. S8: Phase-field out-of-plane strain maps for (a) tweed nanostructures and (b) spinel nano-cuboids

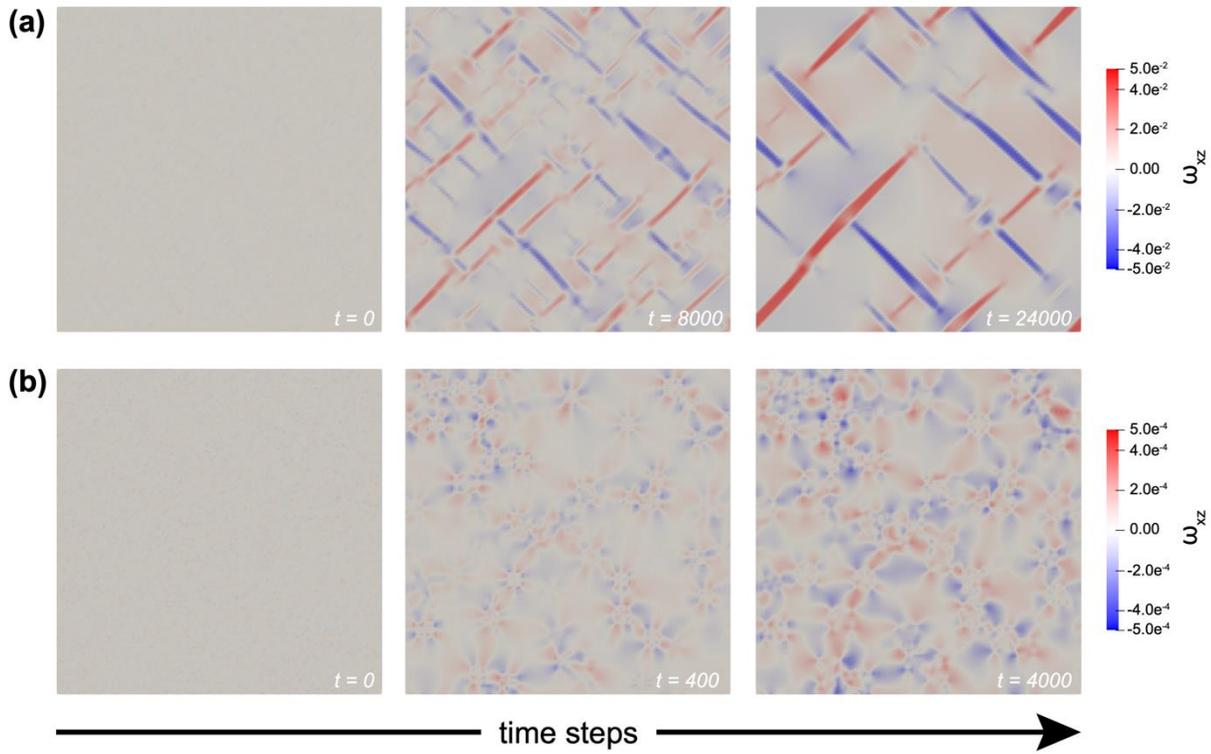

Fig. S9: Phase-field rotational strain maps for (a) tweed nanostructures and (b) spinel nano-cuboids

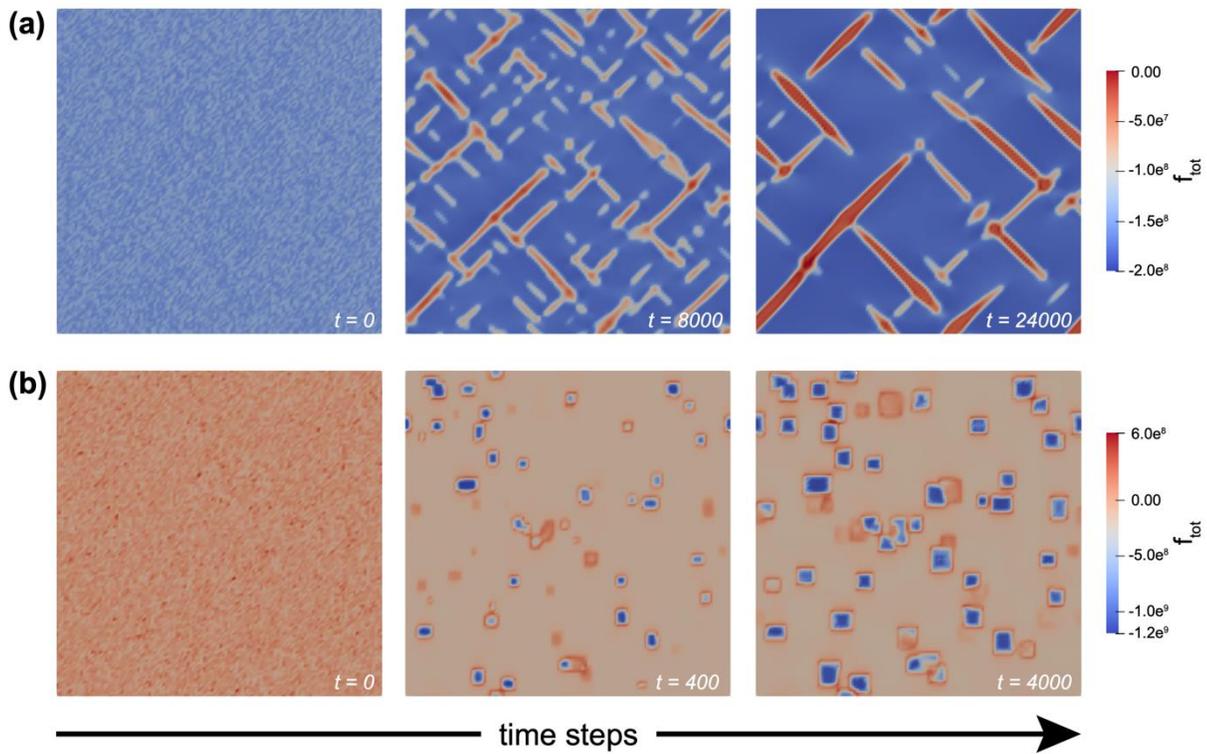

Fig. S10: Phase-field total energy density for (a) tweed nanostructures and (b) spinel nano-cuboids

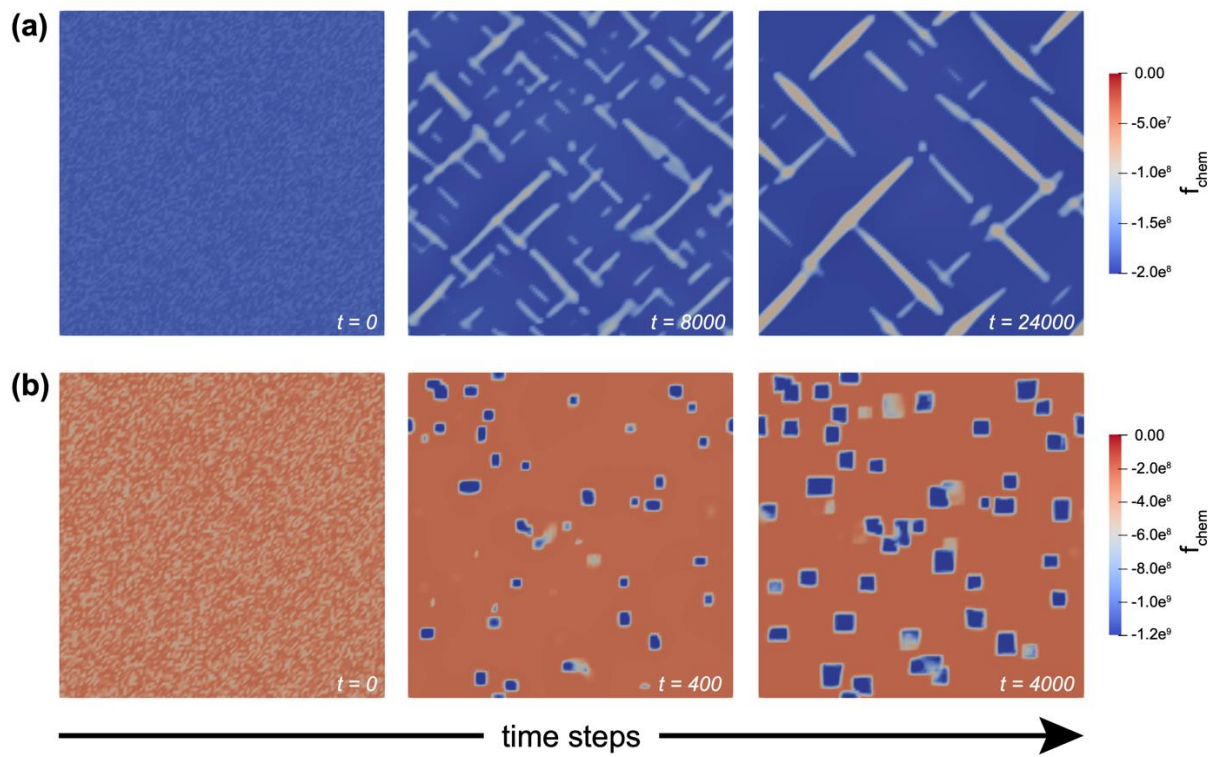

Fig. S11: Phase-field chemical energy density for (a) tweed nanostructures and (b) spinel nano-cuboids

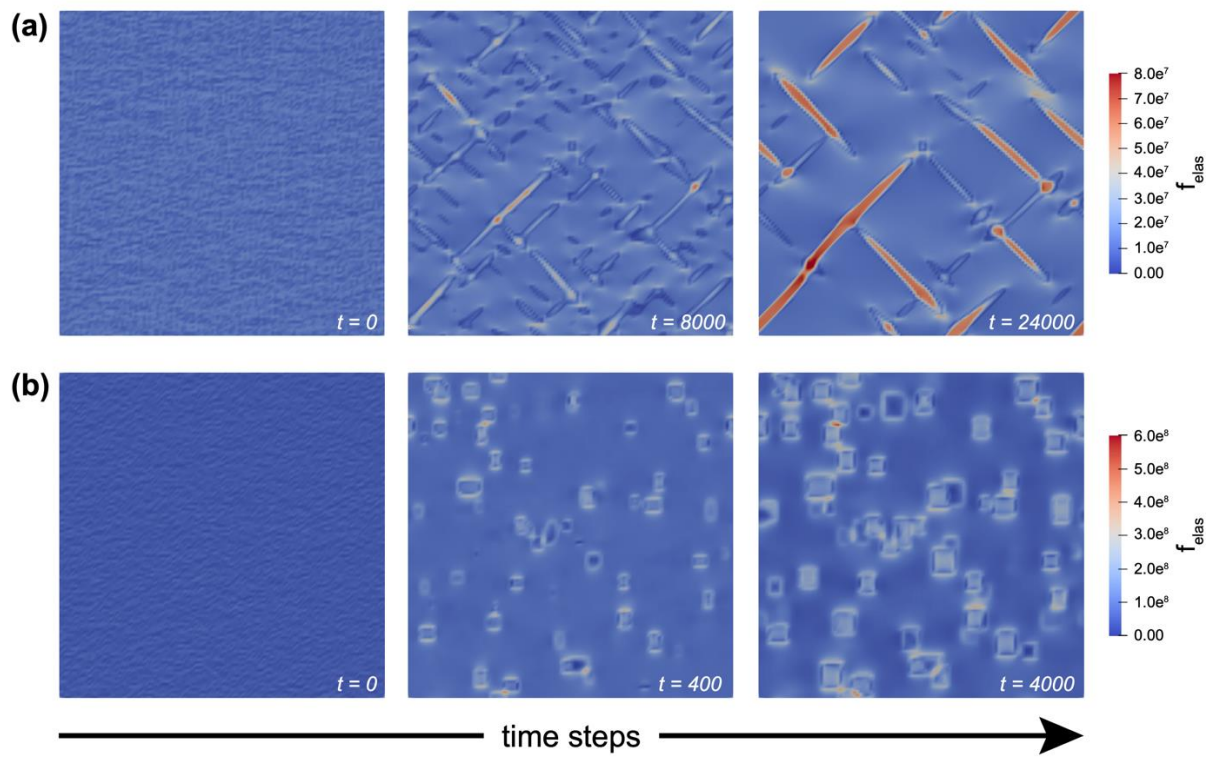

Fig. S12: Phase-field elastic/strain energy density for (a) tweed nanostructures and (b) spinel nano-cuboids

**Supplementary Note 6: Phase-Field Modeling Methodology:**

The phase-field method used to simulate each precipitate phase defines an order parameter $\eta$ to distinguish between the matrix solution phase ($\eta = 0$) and precipitate phase ($\eta = 1$)[3]. The temporal evolutions of the order parameter is carried out by minimizing the free energy $F$ of the system through the Allen-Cahn[4] equation,

$$\frac{\delta \eta}{\delta t} = -L \frac{\delta F}{\delta \eta}$$

where $t$ is the time and $L$ is the linear kinetic parameter, whose value is $1.0 \, \text{m}^3\text{s/J}$ for all simulations in this study. The free energy considers the contributions from the interfacial free energy density $f_{\text{int}}$, elastic strain energy density $f_{\text{elas}}$, and the bulk free energy density $f_{\text{bulk}}$,

$$F = \int [f_{\text{bulk}}(x_i, \eta, T) + f_{\text{int}}(\eta, \nabla \eta) + f_{\text{elas}}(x_i, \eta)] \, dV$$

where $T$ is the thermodynamic temperature and $x_i$ ($i$ = Mg, Ni, Co, Cu, Zn) are the cation concentrations within the rock salt matrix phase. The integration is carried out across the entire film section simulated.

The bulk free energy density describes the chemical contribution to the free energy from both the matrix and precipitate phases. In phase-field simulations, we consider the CuO-rich phase and the spinel precipitate as compounds with fixed compositions, while treating the rock salt matrix as a solid solution, following established work for phase-field with stoichiometric compound[5],

$$f_{\text{bulk}} = \frac{1}{V_m} \{[1 - h(\eta)] \mu^{\text{ss}}(x_i, T) + h(\eta) \mu_{\text{comp}}(T)\}$$

where $V_m$ is the molar volume, $\mu^{\text{ss}}$ is the chemical potential of the rock salt solid solution and is a function of the growth temperature and cation concentrations, $\mu_{\text{comp}}$ is the chemical potential for the stoichiometric compound phase, and $h(\eta) = \eta^3(6\eta^2 - 15\eta + 10)$ is an interpolation function which satisfies $h(0) = 0$, $h(1) = 1$ and $h'(0) = h'(1) = 0$. The concentrations of the cations $x_i^{\text{tot}} = [1 - h(\eta)] x_i + h(\eta) v_i$ are governed by diffusion equations, where +2 cation diffusion coefficients are estimated based on MgO experimental values.

The interfacial energy is described using a combination of a double-well potential and a gradient term,

$$f_{\text{int}} = A\eta^2(1 - \eta)^2 + f_{\text{grad}}(\nabla \eta)$$

where the parameter $A$ is related to the height of the potential well. For CuO precipitate in this study, the anisotropy in interfacial energy density is omitted as elastic anisotropy is estimated to be large enough to govern the formation of a tweed pattern, and the gradient term $f_{\text{grad}}(\nabla \eta)$ is simply formulated as $f_{\text{grad}} = 1/2 \cdot \kappa |\nabla \eta|^2$ where $\kappa$ is a scalar. These two parameters $A$ and $\kappa$ could be determined through the interfacial energy and the thickness of the interface. For spinel morphology, the anisotropic interfacial energy density is playing an active role and is formulated as

$$f_{\text{grad}}(\nabla \eta) = \frac{K_1(\eta_{,1}^6 + \eta_{,2}^6 + \eta_{,3}^6) + K_2[\eta_{,1}^2(\eta_{,2}^4 + \eta_{,3}^4) + \eta_{,2}^2(\eta_{,3}^4 + \eta_{,1}^4) + \eta_{,3}^2(\eta_{,1}^4 + \eta_{,2}^4)] + K_3 \eta_{,1}^2 \eta_{,2}^2 \eta_{,3}^2}{2(\eta_{,1}^2 + \eta_{,2}^2 + \eta_{,3}^2)^2}$$

where $\eta_{,i} := \partial\eta/\partial x_i$ and $\{K_1, K_2, K_3\}$ are parameters determining the interfacial anisotropy. These parameters are estimated from the ratios between low-index surface energies for Co$_3$O$_4$ spinel. The elastic interaction energy is described by a linear elastic model with a stress-free strain introduced by the phase transformation and change in chemical profile,

$$f_{\text{elas}}(x_i, \eta) = \frac{1}{2} C_{ijkl} e_{ij} e_{kl} = \frac{1}{2} C_{ijkl} [\varepsilon_{ij} - \varepsilon_{ij}^0(x_i, \eta)][\varepsilon_{kl} - \varepsilon_{kl}^0(x_i, \eta)]$$

where the elastic strain $e_{ij}$ is determined by the difference between the total strain $\varepsilon_{ij} = \frac{1}{2}\left(\frac{\partial u_i}{\partial x_j} + \frac{\partial u_j}{\partial x_i}\right)$ and the stress-free strain $\varepsilon_{ij}^0(x_i, \eta)$ [6]. More details and modifications on the computational model will be discussed in future publications. Some parameters of note used in phase-field simulations are listed in Table S1.

Table S1: Phase-field simulations parameters.

| Cu-rich phase | | Spinel phase | |
| --- | --- | --- | --- |
| Parameter | Value | Parameter | Value |
| A | $7 \times 10^8$ J/m$^3$ | A | $9 \times 10^9$ J/m$^3$ |
| | | $K_1$ | $7.2 \times 10^{-10}$ J/m |
| κ | $1.05 \times 10^{-10}$ J/m | $K_2$ | $4.1 \times 10^{-9}$ J/m |
| | | $K_3$ | $2.64 \times 10^{-8}$ J/m |

**Supplementary Note 7: Density Functional Theory Methodology:**

Density functional theory (DFT) calculations were performed with the Vienna Ab-initio Simulation Package (VASP)[7] 5.4.1 using the modified Perdew-Burke-Ernzerhof functional for solids (PBEsol)[8] and the standard PAW potentials. DFT+U was used to more properly describe the localized *d*-electrons of Co (4 eV), Ni (5 eV), and Cu (5eV) with the approach of Dudarev[9]. These U values were selected based on sensitivity tests for the binary rock salt oxides that were compared to values from experiment or a hybrid functional (HSE06[10]). A plane-wave cutoff of 700 eV with a 3x3x3 Γ-centered Monkhorst-Pack[11] k-point mesh and Gaussian smearing of 0.05 eV was found to converge the binary oxide total energies to within 1 meV/atom. The magnetic ions (Co, Ni, Cu) were initialized with an antiferromagnetic (AFM) order, with ferromagnetic planes alternating in the [111] direction as found previously for Mg$_{0.2}$Co$_{0.2}$Ni$_{0.2}$Cu$_{0.2}$Zn$_{0.2}$O[12]. Special quasi-random structures[13] (SQSs) of the ternary oxides were generated with the integrated cluster expansion toolkit[14] (ICET). 64-atom 2x2x2 supercells of the rock salt unit cell were used for the ternary oxides and tetragonally distorted rock salt CuO, while the 56-atom spinel unit cell was used for Co$_3$O$_4$. Convergence was assumed when the energy difference between two consecutive self-consistent cycles was less than 10$^{-6}$ eV and structures were relaxed until total force was less than 5 meV/Å. Cubic symmetry was enforced for the ternary oxide SQSs and spinel Co$_3$O$_4$, while the symmetry of CuO was allowed to relax to the tetragonally distorted rock salt structure. Atomic coordinates were allowed to relax in all cases. Elastic stiffness constants were calculated using the stress-strain relationship determined from the relaxed structures. Table 2S lists DFT calculated values.

Table S2: DFT calculated lattice parameters, elastic stiffness values, and Voigt-Reuss-Hill (VRH)[15] bulk ($K_{VRH}$) and shear ($G_{VRH}$) moduli. J14$_{avg}$ is the average of the ten ternary (AA')O SQS values which was used to approximate $Mg_{0.2}Co_{0.2}Ni_{0.2}Cu_{0.2}Zn_{0.2}O$ in the phase-field simulations. As the ternary oxide SQSs and $Co_3O_4$ had cubic symmetry, only the symmetry inequivalent elastic constants ($c_{11}$, $c_{12}$, and $c_{44}$) are reported.

| Compound | Lattice Parameter | $c_{11}$ / $c_{22}$ / $c_{33}$ | $c_{12}$ / $c_{23}$ / $c_{31}$ | $c_{44}$ / $c_{55}$ / $c_{66}$ | $K_{VRH}$ | $G_{VRH}$ |
|---|---|---|---|---|---|---|
| | (Å) | (GPa) | (GPa) | (GPa) | (GPa) | (GPa) |
| Distorted CuO | 3.9144 [a] <br> 5.1330 [c] | 454.3 <br> 454.3 <br> 78.6 | 197.2 <br> 77.0 <br> 77.0 | 32.2 <br> 32.2 <br> 124.2 | 133.1 | 63.5 |
| $Co_3O_4$ | 8.0331 | 301.4 | 186.0 | 103.7 | 224.5 | 82.0 |
| MgCoO | 4.2155 | 307.2 | 114.4 | 110.1 | 178.6 | 104.4 |
| MgCuO | 4.2143 | 281.5 | 127.9 | 104.8 | 179.1 | 92.5 |
| MgNiO | 4.1792 | 352.2 | 104.4 | 116.7 | 187.0 | 119.5 |
| MgZnO | 4.2445 | 280.3 | 124.7 | 113.8 | 176.5 | 97.7 |
| CoCuO | 4.2009 | 308.2 | 148.0 | 71.8 | 201.4 | 75.0 |
| CoNiO | 4.1845 | 355.1 | 127.1 | 81.5 | 203.1 | 93.3 |
| CoZnO | 4.2485 | 285.4 | 144.9 | 79.1 | 191.7 | 75.4 |
| CuNiO | 4.1779 | 332.3 | 140.8 | 78.8 | 204.7 | 85.2 |
| CuZnO | 4.2457 | 277.2 | 150.5 | 75.8 | 192.7 | 70.5 |
| NiZnO | 4.2143 | 325.5 | 136.8 | 85.3 | 199.7 | 88.8 |
| J14$_{avg}$ | 4.2125 | 310.5 | 131.9 | 91.8 | 191.5 | 90.8 |

**Supplementary References:**

(1) Kotsonis, G. N.; Meisenheimer, P. B.; Miao, L.; Roth, J.; Wang, B.; Shafer, P.; Engel-Herbert, R.; Alem, N.; Heron, J. T.; Rost, C. M.; Maria, J.-P. Property and Cation Valence Engineering in Entropy-Stabilized Oxide Thin Films. *Phys. Rev. Mater.* **2020**, *4* (10), 100401. https://doi.org/10.1103/PhysRevMaterials.4.100401.

(2) Meisenheimer, P. B.; Williams, L. D.; Sung, S. H.; Gim, J.; Shafer, P.; Kotsonis, G. N.; Maria, J.-P.; Trassin, M.; Hovden, R.; Kioupakis, E.; Heron, J. T. Magnetic Frustration Control through Tunable Stereochemically Driven Disorder in Entropy-Stabilized Oxides. *Phys. Rev. Mater.* **2019**, *3* (10), 104420. https://doi.org/10.1103/PhysRevMaterials.3.104420.

(3) Chen, L.-Q. Phase-Field Models for Microstructure Evolution. *Annual review of materials research* **2002**, *32* (1), 113–140.

(4) Allen, S. M.; Cahn, J. W. A Microscopic Theory for Antiphase Boundary Motion and Its Application to Antiphase Domain Coarsening. *Acta metallurgica* **1979**, *27* (6), 1085–1095.

(5) Ji, Y.; Chen, L.-Q. Phase-Field Model of Stoichiometric Compounds and Solution Phases. *Acta Materialia* **2022**, *234*, 118007. https://doi.org/10.1016/j.actamat.2022.118007.

(6) Kinoshita, N.; Mura, T. Elastic Fields of Inclusions in Anisotropic Media. *physica status solidi (a)* **1971**, *5* (3), 759–768. https://doi.org/10.1002/pssa.2210050332.

(7) Kresse, G.; Furthmüller, J. Efficient Iterative Schemes for Ab Initio Total-Energy Calculations Using a Plane-Wave Basis Set. *Phys. Rev. B* **1996**, *54* (16), 11169–11186. https://doi.org/10.1103/PhysRevB.54.11169.

(8) Perdew, J. P.; Ruzsinszky, A.; Csonka, G. I.; Vydrov, O. A.; Scuseria, G. E.; Constantin, L. A.; Zhou, X.; Burke, K. Restoring the Density-Gradient Expansion for Exchange in Solids and Surfaces. *Phys. Rev. Lett.* **2008**, *100* (13), 136406. https://doi.org/10.1103/PhysRevLett.100.136406.

(9) Dudarev, S. L.; Botton, G. A.; Savrasov, S. Y.; Humphreys, C. J.; Sutton, A. P. Electron-Energy-Loss Spectra and the Structural Stability of Nickel Oxide: An LSDA+U Study. *Phys. Rev. B* **1998**, *57* (3), 1505–1509. https://doi.org/10.1103/PhysRevB.57.1505.

(10) Krukau, A. V.; Vydrov, O. A.; Izmaylov, A. F.; Scuseria, G. E. Influence of the Exchange Screening Parameter on the Performance of Screened Hybrid Functionals. *The Journal of Chemical Physics* **2006**, *125* (22), 224106. https://doi.org/10.1063/1.2404663.

(11) Monkhorst, H. J.; Pack, J. D. Special Points for Brillouin-Zone Integrations. *Phys. Rev. B* **1976**, *13* (12), 5188–5192. https://doi.org/10.1103/PhysRevB.13.5188.

(12) Zhang, J.; Yan, J.; Calder, S.; Zheng, Q.; McGuire, M. A.; Abernathy, D. L.; Ren, Y.; Lapidus, S. H.; Page, K.; Zheng, H.; Freeland, J. W.; Budai, J. D.; Hermann, R. P. Long-Range Antiferromagnetic Order in a Rocksalt High Entropy Oxide. *Chem. Mater.* **2019**, *31* (10), 3705–3711. https://doi.org/10.1021/acs.chemmater.9b00624.

(13) Zunger, A.; Wei, S.-H.; Ferreira, L. G.; Bernard, J. E. Special Quasirandom Structures. *Phys. Rev. Lett.* **1990**, *65* (3), 353–356. https://doi.org/10.1103/PhysRevLett.65.353.

(14) Ångqvist, M.; Muñoz, W. A.; Rahm, J. M.; Fransson, E.; Durniak, C.; Rozyczko, P.; Rod, T. H.; Erhart, P. ICET – A Python Library for Constructing and Sampling Alloy Cluster Expansions. *Advanced Theory and Simulations* **2019**, *2* (7), 1900015. https://doi.org/10.1002/adts.201900015.

(15) Chung, D.; Buessem, W. The Voigt-reuss-hill Approximation and Elastic Moduli of Polycrystalline MgO, CaF2, β-ZnS, ZnSe, and CdTe. *Journal of Applied Physics* **1967**, *38* (6), 2535–2540.